\documentclass[%
 reprint,
superscriptaddress,
 amsmath,amssymb,longbibliography,
 aps,
prb,
]{revtex4-2}
\pdfoutput=1
\usepackage{ulem}
\usepackage{hyperref}
\usepackage{graphicx}
\usepackage{amsmath}
\usepackage{color}
\usepackage{physics}
\usepackage{dsfont}
\usepackage{comment}
\usepackage{physics}
\usepackage{booktabs}
\usepackage{pifont}
\usepackage[T1]{fontenc}
\usepackage{makecell}
\usepackage{pdfpages} 
\usepackage{pgffor} 
\usepackage{multirow}

\makeatletter
\AtBeginDocument{\let\LS@rot\@undefined}
\makeatother

\def\supplementfilename{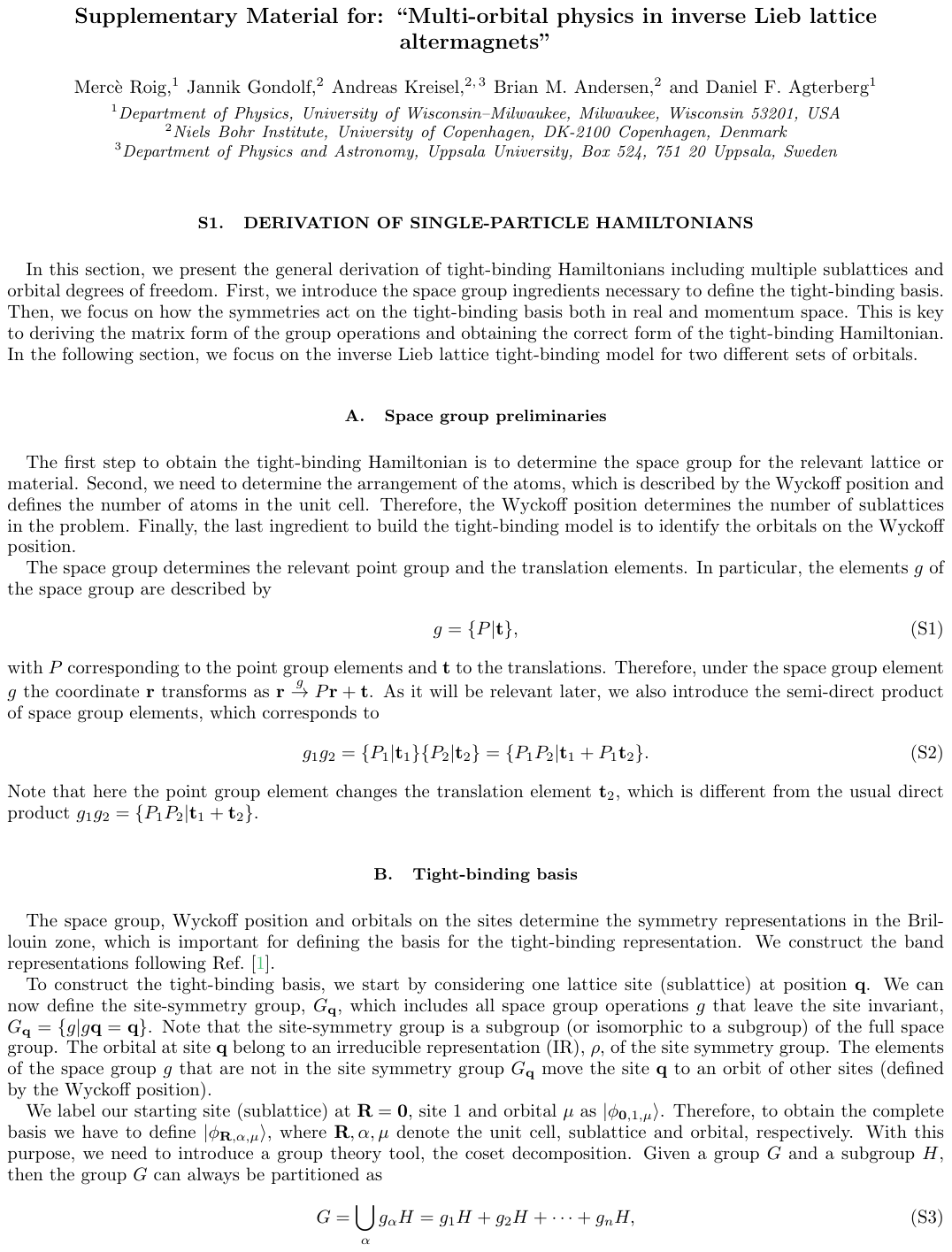}
\def\numbersupplementpages{\the\pdflastximagepages}

\newif\ifarXiv
\arXivtrue 

\definecolor{limegreen}{rgb}{0.2, 0.8, 0.2}
\definecolor{orange}{rgb}{1.0, 0.5, 0.0}

\definecolor{emerald}{rgb}{0.31, 0.78, 0.47}
\definecolor{blue(ncs)}{rgb}{0.0, 0.53, 0.74}

\hypersetup{
     colorlinks=true,
     linkcolor=magenta,
     filecolor=blue,
     citecolor=emerald,      
     urlcolor =blue(ncs),
}

\newcommand{\kv}{{\bf k}}
\newcommand{\qv}{{\bf q}}
\newcommand{\Rv}{{\bf R}}
\newcommand{\Gv}{{\bf G}}
\newcommand{\tv}{{\bf t}}
\newcommand{\rv}{{\bf r}}
\newcommand{\zerov}{{\bf 0}}

\begin{document}
\title{Multi-orbital physics in inverse Lieb lattice altermagnets}

\author{Mercè Roig}
\affiliation{Department of Physics, University of Wisconsin–Milwaukee, Milwaukee, Wisconsin 53201, USA} 

\author{Jannik Gondolf}
\affiliation{Niels Bohr Institute, University of Copenhagen, DK-2100 Copenhagen, Denmark} 

\author{Andreas Kreisel}
\affiliation{Niels Bohr Institute, University of Copenhagen, DK-2100 Copenhagen, Denmark} 
\affiliation{Department of Physics and Astronomy, Uppsala University, Box 524, 751 20 Uppsala, Sweden}

\author{Brian M. Andersen}
\affiliation{Niels Bohr Institute, University of Copenhagen, DK-2100 Copenhagen, Denmark} 

\author{Daniel F. Agterberg}
\affiliation{Department of Physics, University of Wisconsin–Milwaukee, Milwaukee, Wisconsin 53201, USA}


\vskip 1cm

\begin{abstract}
The inverse Lieb lattice has recently emerged as a promising platform for altermagnetism, with several materials with this structure proposed as $d$-wave altermagnetic candidates. Here, we develop a symmetry-based microscopic Hamiltonian for these materials that includes both sublattice and orbital degrees of freedom, going beyond the sublattice-only minimal models that have been extensively used to study such altermagnets. 
We apply these models to examine multi-orbital electron correlation physics in the vanadium oxychalcogenide family altermagnets, which contain dominant $xy$ and $xz/yz$ orbitals character at the Fermi level in the  altermagnetic state. 
We demonstrate that $xy$ orbitals are crucial to stabilize the altermagnetic state observed within a single V$_2$O layer, and altermagnetic order in the $xz/yz$ orbitals is induced through Hund's coupling. Additionally, we show that these multi-orbital models reveal topological regimes in which topological edge states are naturally orbital selective.
\end{abstract}

\maketitle

\section{Introduction}
Altermagnets have recently emerged as a new class of magnetic materials exhibiting unique properties that are promising for spintronics applications~\cite{Smejkal2022Sep,Smejkal2022Dec,Hayami2019Nov,Ma2021May,Gonzalez-Hernandez2021Mar,Shao2021Dec,Bai2024Dec,Song2025Jun}, including nonrelativistic spin-split band structures despite a vanishing net magnetization and a large anomalous Hall effect~\cite{Smejkal2020Jun,Smejkal2022Jun,GonzalezBetancourt2023Jan,Roig2024Oct,Osin2025Nov,Attias2024Sep,Takahashi2025May,Fakhredine2023Sep,Benny2026Feb,McClarty2024Apr,Schiff2025Sep}.
A common class of microscopic models used to understand  the origin and properties of altermagnets is minimal models that are based on the site symmetry of two magnetic atoms within the unit cell~\cite{Roig2024Oct,Antonenko2025Mar}. These include only singly-degenerate electronic orbitals of the site symmetry,  yielding models with four electronic degrees of freedom.  The relative simplicity of these models have allowed a series of insights, including an analysis of the role of impurities and domain walls~\cite{Gondolf2025May,Sorn2025Apr,Sorn2025Dec,Vakili2026Mar}, the origin of the anomalous Hall effect~\cite{Roig2024Oct,Osin2025Nov,Attias2024Sep,Takahashi2025May,Venderbos2025Dec,Sarkar2026Apr}, the discovery of quasi-symmetry and the size of spin-ferromagnetic moments~\cite{Roig2025Jul}, piezomagnetism~\cite{Takahashi2025May,Khodas2026Mar,Bell2026Feb}, the role of topology~\cite{Antonenko2025Mar,Fernandes2024Jan,Jiang2026Feb,Fang2024Sep,Liu2025Oct}, the role of quantum geometry~\cite{Heinsdorf2025}, the physics of altermagnetic spin textures~\cite{Schrade2026Feb,Maiani2026Feb,Dar2026Jul}, and the interplay of superconductivity and altermagnetism~\cite{Wu2025Oct,Rasmussen2026,Radevych2026Apr,Sumita2025Oct,Chakraborty2025Jul,Zou2026Jul,Lu2025Oct}.

In practice, most observed altermagnets include multiple electronic orbitals at each magnetic site and the role of inter-orbital electronic correlations remain unclear. A class of materials for which this is important is the AV$_2$Q$_2$O vanadium oxychalcogenides, where A = K, Rb, Cs and Q = S, Se, Te. In these materials, evidence for room-temperature altermagnetism was reported for RbV$_2$Te$_2$O and KV$_2$Se$_2$O~\cite{Jiang2025Mar,Zhang2025Mar,Hu2026Jan,Wang2025Dec,Yang2026Mar}. While it is believed  that individual V$_2$O layers exhibit altermagnetic order, there is debate as to whether or not these layers are stacked ferromagnetically, leading to 3D altermagnetic order, or antiferromagnetically, leading to antiferromagnetic order~\cite{Sun2025Nov,Thapa2026Feb}.  This stacking is determined by a weak interlayer coupling \cite{Hu2026Jan}, revealing that the intralayer altermagnetic order in this class of materials provides a platform to understand the origin of altermagnetism. Furthermore, it has been found through DFT that these materials exhibit a dominant role of $xy$ and $xz/yz$ orbitals at the Fermi surface~\cite{Jiang2025Mar,Zhang2025Mar}, motivating the need to examine the role of multi-orbital electronic correlations. 

In the single orbital-limit, the vanadium oxychalcogenides are described by the so-called inverse Lieb-lattice model for altermagnetism.  The inverse Lieb lattice model has become a widely used minimal tight-binding model for exploring altermagnetism in two dimensions~\cite{Brekke2023Dec,Antonenko2025Mar,Kaushal2025Oct,Durrnagel2025Jul,Radhakrishnan2026Feb,Fu2025Jul,Wang2026Apr,Xu2025Aug,Wu2025Oct,Huo2026Jan,Jang2026Jun,Cheng2026Feb,Chang2025Aug,Wang2026Jan,Fu2025Dec,Yang2025Nov,Wei2025Feb,Liu2025Dec,Mu2026Apr,deCarvalho2026May,Li2026Jul}, as it incorporates the sublattice degree of freedom of the V atoms, which is crucial for capturing altermagnetic signatures~\cite{Roig2024Oct,Gondolf2025May}.
Here, we generalize this model to go beyond the single-orbital limit to include $xy$ and $xz/yz$ orbitals on the V sites, allowing us to incorporate electronic correlations beyond the standard intraorbital Coulomb repulsion. In addition, while the orthorhombic symmetry of the V site ensures that only singly degenerate orbitals are relevant, suggesting that the single-particle Hamiltonians for these orbitals should adopt a common symmetry-based form, we find that they differ. Specifically, the Hamiltonian for $xy$ orbitals takes the familiar Lieb lattice model form, while that for $xz/yz$ orbitals does not. By examining the susceptibilities and performing self-consistent calculations, we demonstrate that the $xy$ orbitals play a crucial role in stabilizing altermagnetism, while the $xz/yz$ orbitals become altermagnetic only in the presence of Hund's coupling. We further explore the topological regime~\cite{Antonenko2025Mar} of these models, demonstrating the emergence of topological edge states for both sets of orbitals and finding a regime in which the edge states are orbital selective.

\section{Results and Discussion}
\subsection{Single-particle Hamiltonian}
Motivated by the AV$_2$Q$_2$O-family altermagnets, we construct a model with $xy$ and $xz/yz$ orbitals at the V sites of the unit cell. We consider the 2D limit, focusing on a single V$_2$O layer and ignoring the dispersion along the $c$-axis. A key simplifying feature for the single-particle Hamiltonian in this limit is that the $xy$ and the $xz/yz$ orbitals have opposite mirror symmetry under reflection through the $c$-axis ($\{m_z|\zerov\}$). This implies that these orbitals cannot be coupled by any in-plane hopping matrix elements, and hence the single-particle Hamiltonians for these two sets of orbitals are decoupled. 

Key to constructing the single-particle Hamiltonian is the site symmetry of a V atom, which includes all operations of the space group that leave the V site invariant. This site symmetry is isomorphic to the point group $D_{2h}$, which implies that the fundamental building blocks for a basis of the Hamiltonian are singly-degenerate orbitals on the V-sites. The $xy$ orbitals naturally serve as such a basis and we will denote the $xy$ orbitals on the two V sites with Pauli matrices $\tau^{xy}_i$.  The $xz/yz$ orbitals need more careful consideration. Specifically, while these two orbitals are degenerate within the crystal point group $D_{4h}$, they do not remain so in the V site-symmetry group $D_{2h}$. When an $xz$ orbital is chosen as a local basis element for one V site, the corresponding orbital on the other V site is $yz$ (this is fixed by the four-fold rotation symmetry that relates the two sites). Here, we include only one $xz/yz$ pair and we denote this pair on the two V sites with Pauli matrices $\tau^{xz/yz}_i$. This restriction is justified by noting that DFT calculations reveal that the other $xz/yz$ pair is around 700~meV higher in energy~\cite{Jiang2025Mar}.

\begin{figure}[t]
\begin{center}
\includegraphics[angle=0,width=.95\linewidth]{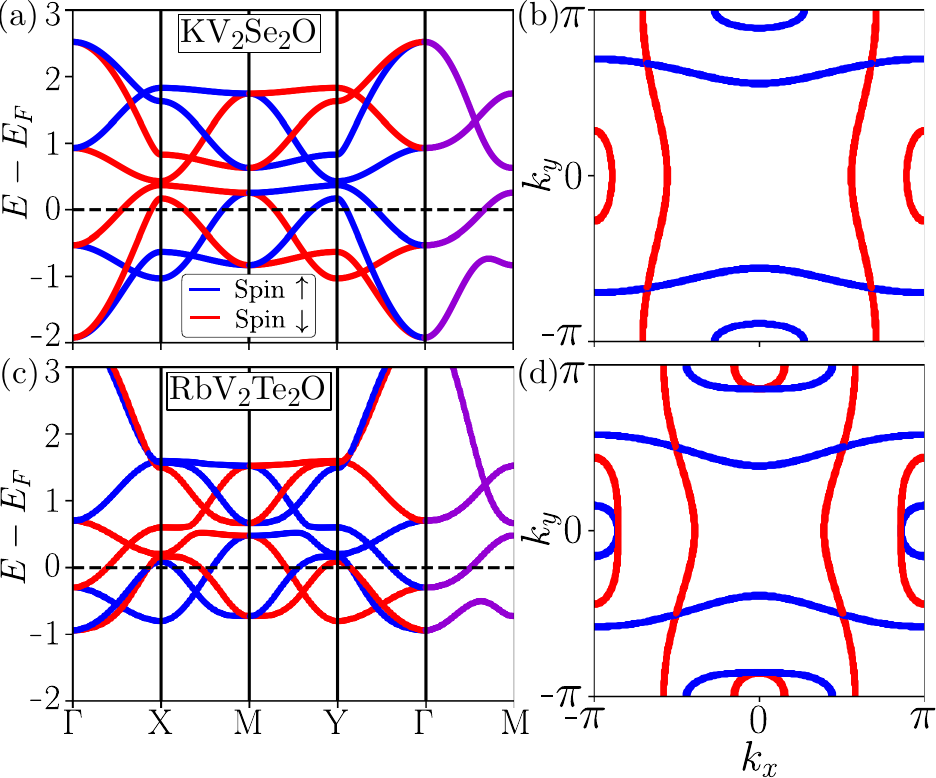}
\caption{Altermagnetic band structures and Fermi surfaces including $xy$ and $xz/yz$ orbitals for (a)-(b) KV$_2$Se$_2$O and (c)-(d) RbV$_2$Te$_2$O obtained from the minimal models in Eqs.~\eqref{eq:general_TBM_AM}-\eqref{eq:tz_hopping}, using the hopping parameters and altermagnetic order $N_z \tau_z \sigma_z$ detailed in Table~\ref{tab:hopping_param}. We emphasize reproducing the Fermi surfaces obtained from DFT calculations~\cite{Jiang2025Mar,Zhang2025Mar} and providing realistic magnetic band splittings.}
\label{fig:TBM_Lieb_lattice}
\end{center}
\end{figure}

As we show below, the tight-binding Hamiltonian for orbital $\mu$ takes the form
\begin{align}
	H^{\mu}_0 (\kv) = \varepsilon_{0,\kv}\tau_0 + t^\mu_{x,\kv} \tau_x^\mu + t_{z,\kv} \tau_z +  \lambda^\mu_{z,\kv}\tau_y^\mu \sigma_z,
	\label{eq:general_TBM_AM}
\end{align}
where $\varepsilon_{0,\kv}$ is the sublattice-independent dispersion, $t_{x,\kv}$ is the hopping between the two sublattices and $\lambda_{\kv}$ is the spin-orbit coupling (SOC). Importantly,  the operators $\tau_{x}^{xy},\tau_y^{xy}$ and $\tau_{x}^{xz/yz},\tau_y^{xz/yz}$ do not share the same symmetry, whereas the corresponding $\tau_0^\mu$ and $\tau_z^\mu$ operators have the same symmetry. This implies that tight-binding Hamiltonians for the $xy$ and the $xz/yz$ orbitals take a different form. Specifically, we find that \begin{align}
	&t^{xy}_{x,\kv} = t_x^{xy} \cos \frac{k_x}{2}\cos\frac{k_y}{2}, \quad \lambda^{xy}_{z,\kv} = \lambda_z^{xy} \sin \frac{k_x}{2}\sin\frac{k_y}{2},\label{eq:hoppings_SOC_xy} \\
    &t^{xz/yz}_{x,\kv}\! {=} t^{xz/yz}_{x}\!\sin \frac{k_x}{2}\sin\frac{k_y}{2}, \lambda^{xz/yz}_{z,\kv} \!{=}\lambda^{xz/yz}_{z}\! \cos \frac{k_x}{2}\cos\frac{k_y}{2}.
	\label{eq:hoppings_SOC_xzyz}
\end{align}
In order to reproduce the Fermi surfaces for RbV$_2$Te$_2$O and KV$_2$Se$_2$O~\cite{Jiang2025Mar,Zhang2025Mar}, we also take
\begin{align}
    \varepsilon_{0,\kv} =& t_1 (\cos k_x + \cos k_y) + t_2 \cos k_x \cos k_y - \mu \label{eq:eps0_hopping}\\
	t_{z,\kv} =& (\cos k_x - \cos k_y) [t_{z,1} + t_{z,2}(\cos k_x +\cos k_y)\nonumber \\
	&+ t_{z,3}\cos k_x \cos k_y].
    \label{eq:tz_hopping}
\end{align}
The additional hopping terms $t_{z,2}$ and $t_{z,3}$ are needed to capture the observed band crossing of opposite spins for RbV$_2$Te$_2$O~\cite{Zhang2025Mar,Hu2026Jan}. Including the altermagnetic order as $H_{\rm AM} = \tau_z \vec{N}\cdot \vec{\sigma}$ allows us to fit the altermagnetic band structures as shown in Fig.~\ref{fig:TBM_Lieb_lattice}, using the hopping parameters detailed in Table~\ref{tab:hopping_param}. Later, we will self-consistently derive the altermagnetic order using our correlated Hamiltonian. 

\begin{table}[t]
\caption{Hopping parameters for $xy$ and $xz/yz$ orbitals (in eV) used to obtain the band structures for KV$_2$Se$_2$O (KVSO) and RbV$_2$Te$_2$O (RVTO) in Fig.~\ref{fig:TBM_Lieb_lattice}, using Eqs.~\eqref{eq:general_TBM_AM}-\eqref{eq:tz_hopping}. The last column corresponds to the Néel order parameter $N_z \tau_z \sigma_z$. The last two rows correspond to the hopping parameters in the topological regime used to obtain Fig.~\ref{fig:edge_states}.}
\begin{tabular}{ccccccccccc} \hline \hline
 & & & & & & & & & & \\[-0.25cm] 
 & Orbitals & $t_1$ & $t_2$ & $t_x$ & $t_{z,1}$ & $t_{z,2}$ & $t_{z,3}$ & $\mu$ & $\lambda_z$ & $N_z$ \\[0.1cm] \hline
 & & & & & & & & & & \\[-0.3cm] 
\multirow{2}{*}{\rotatebox[origin=c]{90}{\scriptsize KVSO}} & $xy$ & 0.1 & -0.2 & 2.1 & 0.2 & 0 & 0 & -0.3 & 0 & 0.73 \\ 
 & $xz/yz$ & -0.2 & 0.1 & 0.15 & 0.35 & 0 & 0 & -0.5 & 0 & 0.73  \\[0.1cm] \hline
 & & & & & & & & & & \\[-0.3cm] 
\multirow{2}{*}{\rotatebox[origin=c]{90}{\scriptsize RVTO}} & $xy$ & 0.35 & -0.08 & 2.2 & 0.025 & -0.2 & 0.005 & -0.75 & 0 & 0.7 \\ 
 & $xz/yz$ & -0.2 & 0.1 & 0.15 & 0.35 & 0 & 0 & -0.5 & 0 & 0.5 \\[0.1cm] \hline
 & & & & & & & & & & \\[-0.3cm] 
\multirow{2}{*}{\rotatebox[origin=c]{90}{\scriptsize \makecell{KVSO\\(topo.)}}} & $xy$ & 0.1 & -0.2 & 2.1 & 0.3 & 0 & 0  & -0.1 & 0.25 & 0.45 \\ 
 & $xz/yz$ & -0.2 & 0.05 & 0.45 & 0.35 & 0 & 0 & -0.6 & 0.25 & 0.45 \\[0.1cm] 
 \hline \hline
\end{tabular}
\label{tab:hopping_param}
\end{table}

To understand the origin of the differences between the Hamiltonians for $xy$ and $xz/yz$ orbitals, it is useful to develop the theory more carefully. Formally, these materials belong to the symmorphic space group 123 ($P_{4/mmm}$), and the Wyckoff position for the V site is 2f, corresponding to the positions $(1/2,0)$ and $(0,1/2)$ in a 2D model. 
The generators of the space group are $\{E|\tv_1\}$, $\{E|\tv_2\}$, $\{C_{4z}|\zerov\}$, $\{I|\zerov\}$, $\{m_x|\zerov\}$ and $\{m_z|\zerov\}$, with $E$, $C_{4z}$, $I$, $m_x$ and $m_z$ denoting the identity, four-fold rotation, inversion and mirrors across the $x$ and $z$ axis, respectively, and $\tv_1$ and $\tv_2$ denoting the generators of the translation group. 
Since the two sublattice sites are related by $\{C_{4z}|\zerov\}$, it is useful to write the space group as
\begin{equation}
    P_{4/mmm} = \{E|\zerov\} P_{mmm} + \{C_{4z}| \zerov \} P_{mmm},
\end{equation}
where we introduce the coset representatives 
$g_1 = \{E|\zerov \}$ and $g_2 = \{C_{4z}| \zerov \}$ for the two sublattices $\alpha=\{1,2\}$ in the unit cell; see the Supplementary Material (SM)~\cite{Supplementary}.
This allows us to define our tight-binding basis as
\begin{equation}
	\ket{\phi_{\Rv, \alpha, \mu}} \equiv \{ E| \Rv\} g_\alpha \ket{\phi_{\zerov,1,\mu}},
    \label{eq:tight_binding_basis}
\end{equation}
with $\{ E| \Rv\}$ denoting the translation element between different unit cells.
Under a space group element $h=\{P|\tv\}$, this basis transforms as~\cite{Cano2018Jan,Supplementary}
\begin{equation}
    h \ket{\phi_{\Rv,\alpha,\mu}} = [U_h]_{\beta \nu, \alpha \mu} \ket{\phi_{\Rv',\beta,\nu}},
    \label{eq:symmetries_real_space}
\end{equation}
where the transformation matrix $[U_h]_{\beta \nu, \alpha \mu}$ is independent of $\Rv$ and $\Rv'$. 

Using the previous transformation rule, it is possible to understand the origin of the difference in the two Hamiltonians. In particular, defining 
 $\ket{\phi_{\zerov,2,\mu}} \equiv \{C_4|\zerov \}  \ket{\phi_{\zerov,1,\mu}}$, we can find the transformation matrix $[U_h]_{\beta \nu, \alpha \mu}$ for the generator $\{C_4|\zerov \}$  through $\{C_4|\zerov \}  \ket{\phi_{0,2,\mu}}$, obtaining 
\begin{eqnarray}
    \{C_4|\zerov \}  \ket{\phi_{\zerov,2,\mu}} && = \{C_4|\zerov \}^2  \ket{\phi_{\zerov,1,\mu}} \nonumber \\ && = \{E|0,-1\}\{C_{2z}|0,1\}\ket{\phi_{\zerov,1,\mu}} \nonumber\\  && = \{E|0,-1\}\rho^\mu (C_{2z}) \ket{\phi_{\zerov,1,\mu}}.
\end{eqnarray}
Here, $\{C_{2z}|0,1\}$ is in the site-symmetry group of $\ket{\phi_{\zerov,1,\mu}}$ and hence is represented by its character $\rho^\mu (C_{2z})$. 
This implies $[U_h]_{\beta \nu, \alpha \mu}$ for $\{C_4|\zerov\}$ is given by the $2\cross 2$ matrix in sublattice space
\begin{equation}
 U^{\mu}_{ \{C_4|\zerov \}}=\begin{pmatrix}
        0 & 1 \\
        \rho^\mu (C_{2z}) & 0 
    \end{pmatrix}.
\end{equation}
The orbital considered now becomes crucial, since $\rho^{xy} (C_{2z}) = +1$, while $\rho^{xz/yz} (C_{2z}) = -1$. This difference underlies the distinct forms of the tight-binding Hamiltonian. 

We choose as our Fourier basis $\ket{\varphi_{\kv,\alpha,\nu}} = \frac{1}{\sqrt{N}}\sum_\kv e^{i \kv \cdot(\Rv + \rv_\alpha)} \ket{\phi_{\Rv,\alpha,\nu}}$, which implies that the tight-binding Hamiltonian transforms as \begin{equation}
    H (\kv) = U_g H (P^{-1}\kv) U_g^{-1},
    \label{eq:TB_symmetries}
\end{equation}
where $P$ corresponds to the rotation for the space group operation $g$, see the SM~\cite{Supplementary}.
Consequently, the symmetries of the tight-binding Hamiltonian are determined by the momentum-independent matrices $U_g$. However, 
 the Hamiltonian is not periodic in the Brillouin Zone (BZ), as it satisfies 
\begin{equation}
    H_{\alpha \mu, \beta \nu} (\kv + \Gv) = e^{i \Gv \cdot (\rv_\beta-\rv_\alpha)} H_{\alpha \mu,\beta \nu} (\kv).
    \label{eq:TB_periodicity}
\end{equation}
Using Eqs.~(\ref{eq:TB_symmetries}) and (\ref{eq:TB_periodicity}), it is possible to derive the tight-binding form given in Eqs.~\eqref{eq:hoppings_SOC_xy}-\eqref{eq:tz_hopping} (see the SM for more details~\cite{Supplementary}).

\subsection{Stabilization of altermagnetism}
In two dimensions, the single-particle Hamiltonians  for $xy$ and $xz/yz$ orbitals are uncoupled, as the mirror symmetry $\{m_z|\zerov\}$ prevents the hopping. As a consequence, the coupling between the two models can only arise from electron interactions or SOC.
At this point, we consider the uncoupled single-particle Hamiltonians from the previous sections together with electron interactions. This is expected to have the largest effect; the case of the SOC is discussed in the SM~\cite{Supplementary}.
Thus, we consider coupling of the orbitals through multi-orbital onsite Coulomb repulsion, including Hubbard repulsion $U'$ and Hund's coupling $J_H$, and employ the standard assumption $U'=U-2J_H$~\cite{Georges2013Dec}.

The full Hubbard-Kanamori Hamiltonian reads $H = \sum_\mu H_0^\mu + H_\mathrm{int}$, including the non-interacting terms in Eq.~\eqref{eq:general_TBM_AM} and the intra- and inter-orbital interactions
\begin{align}
    H_{\rm int} & = U \sum_{i,\mu} n_{i\mu\uparrow} n_{i\mu\downarrow} \nonumber \\
    &\hphantom{=} + \!\sum_{i,\mu\neq\mu'}  \Bigg[ \frac{1}{2} \sum_{\sigma \sigma'} \Big( U'c^\dagger_{i\mu' \sigma'} c^\dagger_{i\mu \sigma} c_{i\mu \sigma} c_{i\mu' \sigma'}\nonumber \\
    &\hphantom{=}-\!J_\mathrm{H} c^\dagger_{i\mu' \sigma'} c^\dagger_{i\mu \sigma} c_{i\mu \sigma'} c_{i\mu' \sigma}\!\Big)
    \!+\!J' c_{i\mu\uparrow}^\dagger c_{i\mu\downarrow}^\dagger c_{i\mu'\downarrow} c_{i\mu'\uparrow} \Bigg],
     \label{eq:int_intraorb}
\end{align}
where $i$ is the site index and $\mu = \{ xy, xz/yz\}$ denotes the orbital. The intra-orbital term in Eq.~(\ref{eq:int_intraorb}) has been previously used to demonstrate that altermagnetism can be stabilized in microscopic models~\cite{Roig2025Jul,Yu2025Mar,Maier2023Sep}.
To analyze the leading instabilities, we use both selfconsistent Hartree-Fock calculations and the random phase approximation (RPA), see the SM~\cite{Supplementary}.

To gain insight into the instabilities of these models, we  first examine the bare susceptibilities for these uncoupled single-particle Hamiltonians, and we later include the correlations. Thus, we obtain an analytic expression for the bare susceptibility in the ferromagnetic ($\chi_0^{\rm FM}$) and altermagnetic ($\chi_0^{\rm AM}$) channels~\cite{Supplementary,Roig2024Oct}. For a general momentum $\qv$ and Matsubara frequency $iq_n$, the difference between the two susceptibilities corresponds to
\begin{multline}
    \big[\chi_0^{\rm AM}- \chi_0^{\rm FM}\big](\qv,iq_n) \\
    {=} \sum_{\kv}\frac{t_{x,\kv} t_{x,\kv+\qv}}{E^+_\kv E^+_{\kv+\qv}} [f_{++} {+} f_{--} {-} (f_{+-} {+} f_{-+})],
    \label{eq:bare_suscept}
\end{multline}
where we have defined the Lindhard term $f_{ab} = \frac{f(E_\kv^a) - f(E_{\kv+\qv}^b)}{iq_n + E_\kv^a - E_{\kv+\qv}^b}$ and $E^\pm_{\kv} =\varepsilon_{0,\kv} \pm \sqrt{t_{x,\kv}^2+t_{z,\kv}^2}$.
This expression highlights the key role of the nearest-neighbor hopping in stabilizing altermagnetism, since for a vanishing $t_{x,\kv}$ the two channels become degenerate.
If the altermagnetic susceptibility is larger than the ferromagnetic susceptibility, it indicates that the altermagnetic phase is favored over ferromagnetism or spin-density wave order.
Note that both altermagnetism and ferromagnetism are $\qv=0$ ordered states.

\begin{figure}[t]
\begin{center}
\includegraphics[angle=0,width=0.9\linewidth]{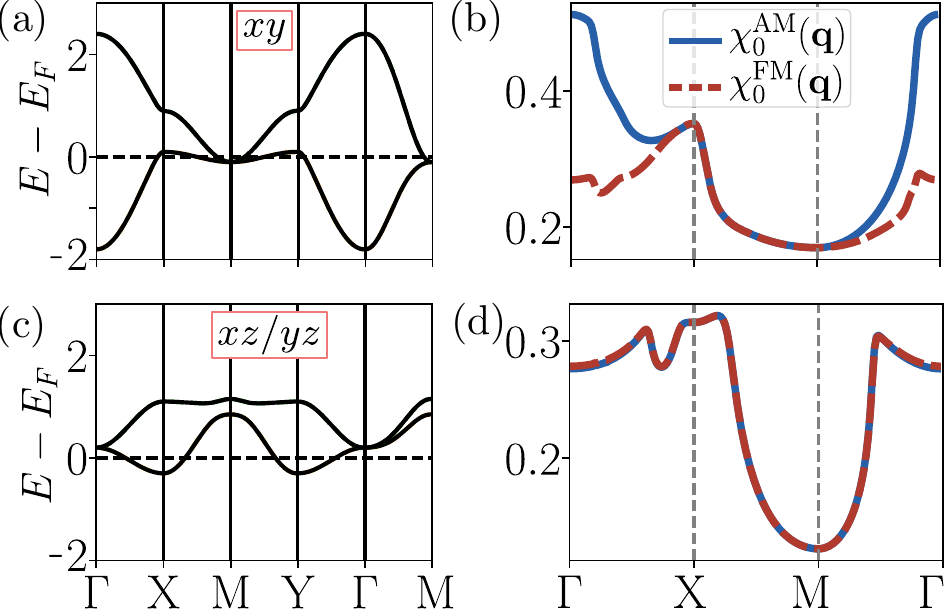}
\caption{Normal state band structures and bare susceptibilities for (a)-(b) $xy$ and (c)-(d) $xz/yz$ orbitals for KV$_2$Se$_2$O. The hopping parameters are detailed in Table~\ref{tab:hopping_param}.}
\label{fig:suscept_Lieb_lattice}
\end{center}
\end{figure}

In Fig.~\ref{fig:suscept_Lieb_lattice}, we show the bare susceptibilities for the normal-state bands for $xy$ and $xz/yz$ orbitals shown in Fig.~\ref{fig:TBM_Lieb_lattice}(a). Notably, the $xy$ orbitals exhibit stronger altermagnetic correlations at $\qv=0$, as seen from the bare susceptibilities, while for the $xz/yz$ orbitals the ferromagnetic and altermagnetic channels are nearly degenerate. This is a consequence of the smaller nearest-neighbor hopping for the $xz/yz$ orbitals, as we demonstrate in the following section by examining in more detail the parameter range that favors altermagnetic correlations for both sets of orbitals. Furthermore, these bare susceptibilities indicate that, while the leading magnetic instability for $xy$ orbitals is altermagnetic, this is not the case for the $xz/yz$ orbitals. These orbitals prefer to order in an incommensurate spin-density wave (SDW) phase.
Similar results are obtained for the model relevant to RbV$_2$Te$_2$O shown in Fig.~\ref{fig:TBM_Lieb_lattice}(c). The RPA treatment of the intra-orbital Hubbard interaction is discussed in the SM for both orbitals~\cite{Supplementary}.

Next we consider the full multi-orbital interaction via Eq.~\eqref{eq:int_intraorb}.
We solve the Hartree-Fock equations for the order parameter selfconsistently, considering mean-fields of the form $\langle c_{\mu \sigma}^\dagger c_{\mu \sigma'} \rangle$, with $\mu \in \{ xy, xz/yz\}$, while neglecting inter-orbital mean fields. This makes the pair hopping term ($J'$) vanish, see the SM~\cite{Supplementary}.
The tight-binding models are obtained from fits to DFT band structures, and therefore faithfully capture the relative filling of the orbitals. Interactions driving a charge transfer between orbitals are already accounted for, and we introduce a crystal-field splitting $\Delta$ in addition to the chemical potential $\tilde\mu$ as Lagrange multiplier to fix the electron density $\langle n_\mu \rangle = \langle n_{\uparrow \mu} + n_{\downarrow \mu} \rangle$ on each of the orbitals
\footnote{Since $U'$ couples only to the orbital densities $n_\mu$ and they are fixed, the relevant interaction for the coupling is Hund's pairing. In particular, the case $J_H=0$ corresponds to fully uncoupled orbitals}.
In order to compare the stability of different ordered phases, we calculate the Helmholtz free energy $F=\Omega + \tilde\mu \langle n \rangle + \Delta(\langle n_{xy} \rangle - \langle n_{xz/yz} \rangle)$, where $\Omega$ is the grand canonical potential and $\tilde\mu$ and $\Delta$ are the self-consistently determined chemical potential and crystal field splitting restricted by the orbitals' fillings.

\begin{figure}[t]
\begin{center}
\includegraphics[width=.95\linewidth]{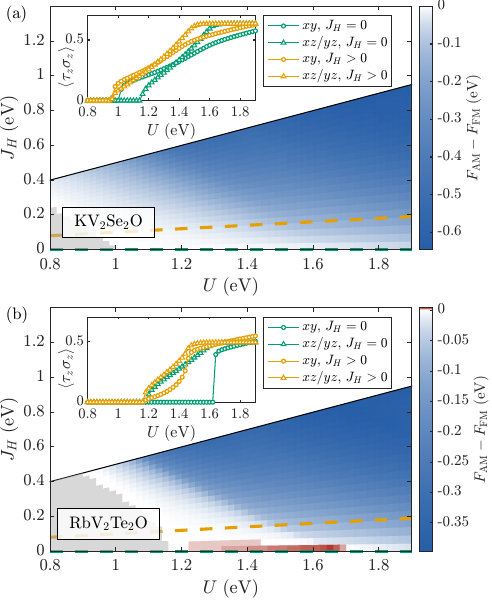}
\caption{Difference in free energy of AM and FM  states for (a) KV$_2$Se$_2$O and (b) RbV$_2$Te$_2$O in $U$–$J_H$ phase space. Blue (red) color shows regions where altermagnetism (ferromagnetism) is favorable. Regions shaded gray represent the nonmagnetic state. The insets show the order parameters as a function of $U$ for $J_H=0$ (green) and $J_H = 0.1U$ (yellow). The $xy$ and $xz/yz$ orbital are shown with circle and triangle markers, respectively. The introduction of finite Hund's coupling results in the concurrent ordering of both orbitals and enhancement of the magnetic order parameter. Calculations are performed at temperature $T=0.02$ eV. 
}
\label{fig:Hunds_pairing}
\end{center}
\end{figure}

As seen in Fig.~\ref{fig:Hunds_pairing}, Hund's coupling can significantly enhance altermagnetism while inducing the simultaneous onset of magnetic order by coupling the orbitals and driving altermagnetism from the $xy$ orbitals to the $xz/yz$ orbitals. In the case of RbV$_2$Te$_2$O, ferromagnetic order is favorable in the absence of Hund's coupling due to the tendency of the $xz/yz$ orbital to order ferromagnetically and a lower critical $U$ for that orbital. Even in that case, however, the $xy$ orbitals drive the whole system to become an altermagnet via the Hund's coupling.
These general findings are robust to the inclusion of relativistic SOC, see the SM~\cite{Supplementary}.

\subsection{Phase diagrams}
In the previous section, we demonstrated that for the parameter set used to obtain Fig.~\ref{fig:suscept_Lieb_lattice}, the leading instability for $xy$ orbitals is altermagnetic, while that for $xz,yz$ orbitals is not. Consequently, Hund's rule coupling between the two orbital sets was needed to drive the $xz/yz$ orbitals altermagnetic. It is reasonable to ask if this result is robust to changes in the parameters used. Here, in order to understand what favors an altermagnetic instability, we examine a broader parameter regime.
For concreteness, we consider the model for $xy$ orbitals and include the hopping parameters $t_1$, $t_x$, $t_{z,1}$ and the chemical potential $\mu$, and set $t_2 = 0$. We fix $t_{z,1}\equiv t_z = 0.2$ to set the energy scale. Note that the hopping $t^{xz/yz}_{x,\kv}$ and the SOC $\lambda^{xz/yz}_{z,\kv}$ in Eq.~\eqref{eq:hoppings_SOC_xzyz} for the $xz/yz$ orbitals correspond to a $(\pi,\pi)$ shift of $t^{xy}_{x,\kv}$ and $\lambda^{xy}_{z,\kv}$ in Eq.~\eqref{eq:hoppings_SOC_xy} for the $xy$ orbitals. This shift also changes the sign of the hopping parameters $t_1$ and $t_z$. Notably, the sign of $t_z$ does not affect the eigenvalues and therefore does not change the phase diagram. 
As a consequence, the phase diagram for the $xz/yz$ orbitals can be obtained from that for the $xy$ orbitals by reversing the sign of $t_1$ or, equivalently, by taking $\mu \rightarrow -\mu$.

\begin{figure}[t]
\begin{center}
\includegraphics[angle=0,width=.95\linewidth]{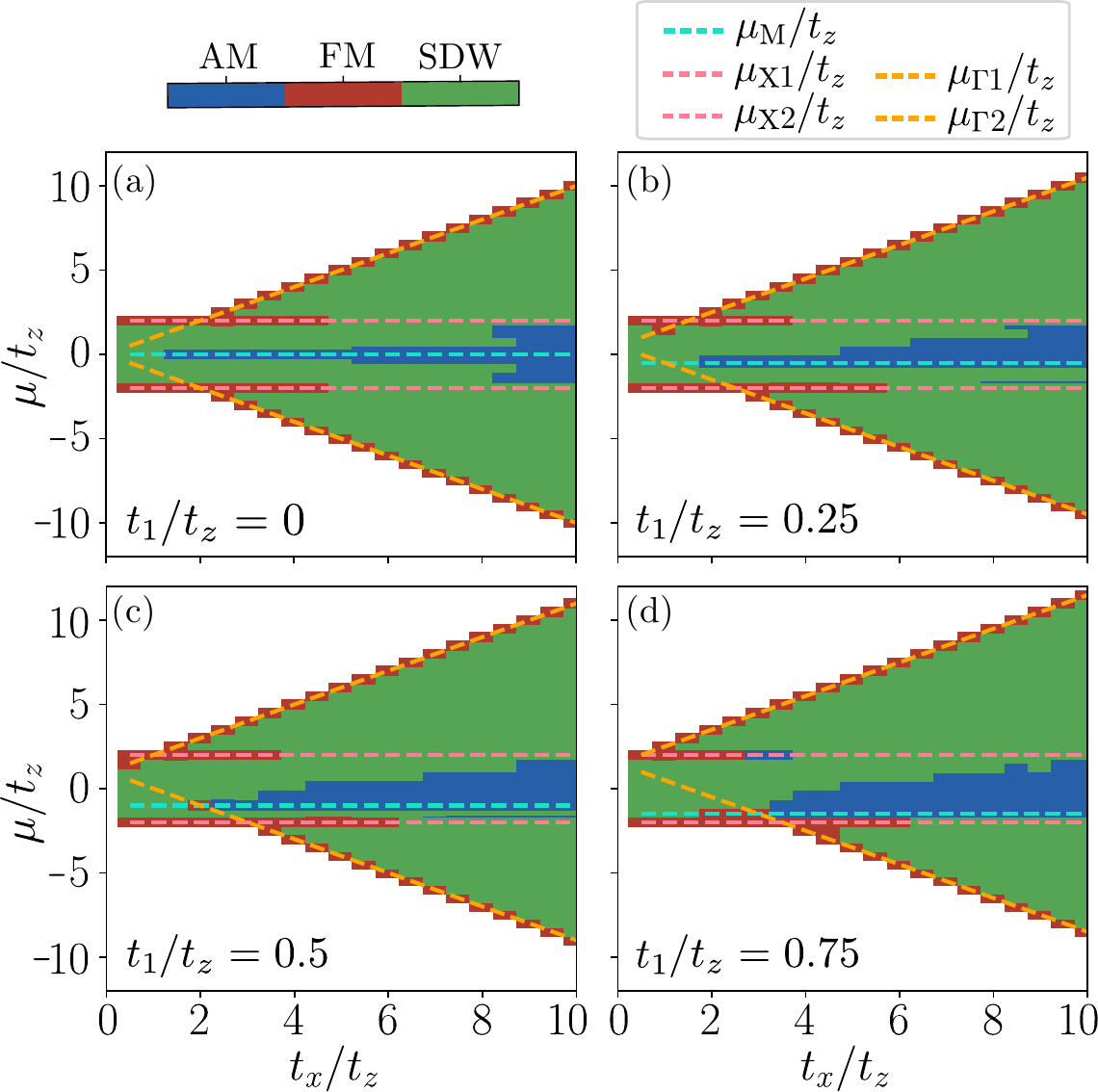}
\caption{Phase diagrams obtained from the bare FM and AM susceptibilities, considering the model for $xy$ orbitals in Eq.~\eqref{eq:hoppings_SOC_xy} and Eqs.~\eqref{eq:eps0_hopping}-\eqref{eq:tz_hopping}, with $t_z = 0.2$. The different phases correspond to altermagnet (AM), ferromagnet (FM), or incommensurate spin-density wave (SDW). The blue, green, and orange dashed lines indicate the values of the chemical potential $\mu/t_z$ at which the M, X, and $\Gamma$ points are at the Fermi level, respectively.}
\label{fig:phase_diagrams}
\end{center}
\end{figure}

To determine the preferred phase, we analyze the bare ferromagnetic ($\chi_0^{\rm FM}(\qv)$) and altermagnetic ($\chi_0^{\rm AM}(\qv)$) susceptibilities, see the SM for the analytic expressions~\cite{Supplementary}. 
If the susceptibility peak occurs at $\qv=0$, we identify the phase as altermagnetic (ferromagnetic) if $\chi_0^{\rm AM}(\mathbf{0})>\chi_0^{\rm FM}(\mathbf{0})$ ($\chi_0^{\rm AM}(\mathbf{0})<\chi_0^{\rm FM}(\mathbf{0})$). If instead the largest peak is at $\qv \neq 0$, then this corresponds to a finite-momentum, or incommensurate, SDW phase.
In Fig.~\ref{fig:phase_diagrams}, we show the phase diagram over a wide range of parameters. In addition, we also indicate by dashed lines when the M, X and $\Gamma$ points are at the Fermi level. Notably, we can identify the key role of van Hove singularities (vHS) in stabilizing altermagnetism. In particular, for a sufficiently large hopping $t_x$, the vHS at the M point helps stabilizing the altermagnetic phase even when the chemical potential lies above this point.

We now consider the model relevant to KV$_2$Se$_2$O shown in Fig.~\ref{fig:suscept_Lieb_lattice} with the hopping parameters detailed in Table~\ref{tab:hopping_param}. In this case, for $xy$ and $xz/yz$ orbitals, we find $t_1/t_z \approx 0.5$, which corresponds to Fig.~\ref{fig:phase_diagrams}(c). For $xz/yz$ orbitals, the nearest-neighbor hopping is small, $t_x/ t_z \approx 0.4$, and, in this regime, altermagnetism is never the stable ground state. This differs for the $xy$ orbitals, for which $t_x/ t_z \approx 10$ and altermagnetism is a viable ground state. This supports the conclusion that Hund's rule is required to drive the $xz/yz$ orbitals altermagnetic. 

\begin{figure}[t]
\begin{center}
\includegraphics[angle=0,width=.95\linewidth]{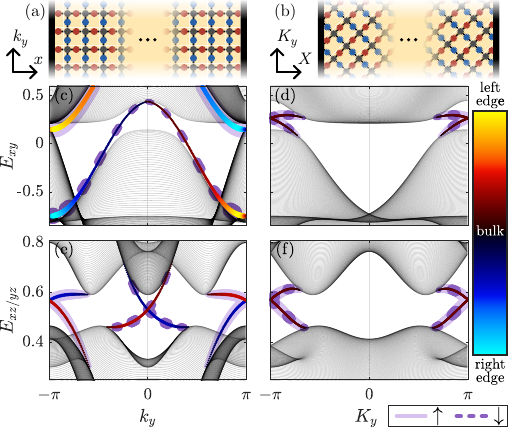}
\caption{Strip geometry of the inverse Lieb lattice with vertical (a) and diagonal (b) termination. The spectra for KV$_2$Se$_2$O in the topological regime for $N_z=0.45$ are shown for the vertical orientation in the left column and for the diagonal orientation in the right column, respectively. (c,d) show the $xy$ orbital and (e,f) the $xz/yz$ orbital, using Eqs.~\eqref{eq:general_TBM_AM}-\eqref{eq:tz_hopping} and the hopping parameters detailed in Table~\ref{tab:hopping_param}. The color of the topological states indicate their amount of localization at the edges and the light purple (dashed dark purple) shading indicates up (down) polarization of the states.
}
\label{fig:edge_states}
\end{center}
\end{figure}

\subsection{Topological edge states}
We have also considered a set of parameters for the altermagnetic band structure for KV$_2$Se$_2$O in a regime where nontrivial topology can emerge, see Table~\ref{tab:hopping_param}.
Notably, the different form of the tight-binding model for the two sets of orbitals in Eqs.~\eqref{eq:hoppings_SOC_xy}-\eqref{eq:hoppings_SOC_xzyz} yields band crossings in different directions in the absence of SOC.
For $xy$ orbitals, the spectrum exhibits Dirac crossing between same spin bands along M-X and M-Y lines protected by the mirror symmetries $m_x$ and $m_y$, respectively. In the presence of SOC, the previous mirror symmetries are no longer preserved, and the spectrum becomes gapped~\cite{Antonenko2025Mar}. In contrast, for $xz/yz$ orbitals, the Dirac crossing maps to the $\Gamma$-X and $\Gamma$-Y directions, which also become gapped when including SOC.
Hence, in the presence of a small altermagnetic moment $N_z \tau_z \sigma_z$, the nontrivial topology allows the emergence of edge states, which disappear as the moment $N_z$ is increased and the bands transition to a trivial phase.

We consider a system with semi-open boundary conditions and surface termination in the $x$ and $X=x-y$ directions, and we calculate the eigenstates of this strip geometry.
In Fig.~\ref{fig:edge_states}, we show the topological edge states obtained from both sets of orbitals for the different lattice terminations. In the vertical geometry, the edge states are spin polarized and propagate in opposite directions at the two opposite edges. In contrast, in the diagonal geometry, the topological states are spin degenerate.
In addition, our selfconsistent calculations reveal the emergence of orbital-selective topological states, as shown in the SM~\cite{Supplementary}. We find that the transition to the topological state, which requires $|N_z/2 t_{z,1}|<1$, occurs at a different $U$ for $xy$ orbitals compared to $xz/yz$ orbitals \footnote{This condition for the transition applies to the model for KV$_2$Se$_2$O. More generally, $\abs{N_z/\max(t_{z,\kv})}<1$ in the topological regime.}. This gives rise to an effective Néel order of different magnitude for the two orbitals.
Therefore, there exists a temperature regime in which the topological states are present only for one orbital component, and are thus orbital selective.

\section{Conclusions}
We have extended minimal models for altermagnetism to include multiple orbital degrees of freedom and applied these models to the vanadium oxychalcogenides family. Our susceptibilities and selfconsistent calculations have revealed the important role of $xy$ orbitals in stabilizing altermagnetism, and the crucial role of Hund's pairing in developing altermagnetism for $xz/yz$ orbitals. Finally, we have shown that these models support orbital-selective topological edge states.

\begin{acknowledgments}
We thank Yue Yu for useful discussions. Work at UWM was supported by National Science Foundation
Grant No. DMREF 2323857 and Simons Foundation Grant
No. SFI-MPS-NFS-00006741-02. J.G. acknowledges support from the Independent Research Fund Denmark, Grant No. 3103-00008B.
B.M.A. acknowledges support from the Independent Research Fund Denmark Grant No. 5241-00007B. A.K. acknowledges support by the Danish National Committee for Research Infrastructure (NUFI) through the ESS-Lighthouse Q-MAT. 
\end{acknowledgments}

\bibliography{AM_Lieb}

\ifarXiv
    

\section*{Supplementary material}

# Supplementary Material for: "Multi-orbital physics in inverse Lieb lattice altermagnets"

Mercè Roig,[1] Jannik Gondolf,[2] Andreas Kreisel,[2,3] Brian M. Andersen,[2] and Daniel F. Agterberg[1]

[1]*Department of Physics, University of Wisconsin–Milwaukee, Milwaukee, Wisconsin 53201, USA*
[2]*Niels Bohr Institute, University of Copenhagen, DK-2100 Copenhagen, Denmark*
[3]*Department of Physics and Astronomy, Uppsala University, Box 524, 751 20 Uppsala, Sweden*

## S1. DERIVATION OF SINGLE-PARTICLE HAMILTONIANS

In this section, we present the general derivation of tight-binding Hamiltonians including multiple sublattices and orbital degrees of freedom. First, we introduce the space group ingredients necessary to define the tight-binding basis. Then, we focus on how the symmetries act on the tight-binding basis both in real and momentum space. This is key to deriving the matrix form of the group operations and obtaining the correct form of the tight-binding Hamiltonian. In the following section, we focus on the inverse Lieb lattice tight-binding model for two different sets of orbitals.

### A. Space group preliminaries

The first step to obtain the tight-binding Hamiltonian is to determine the space group for the relevant lattice or material. Second, we need to determine the arrangement of the atoms, which is described by the Wyckoff position and defines the number of atoms in the unit cell. Therefore, the Wyckoff position determines the number of sublattices in the problem. Finally, the last ingredient to build the tight-binding model is to identify the orbitals on the Wyckoff position.

The space group determines the relevant point group and the translation elements. In particular, the elements $g$ of the space group are described by

$$g = \{P|\mathbf{t}\}, \tag{S1}$$

with $P$ corresponding to the point group elements and $\mathbf{t}$ to the translations. Therefore, under the space group element $g$ the coordinate $\mathbf{r}$ transforms as $\mathbf{r} \xrightarrow{g} P\mathbf{r} + \mathbf{t}$. As it will be relevant later, we also introduce the semi-direct product of space group elements, which corresponds to

$$g_1 g_2 = \{P_1|\mathbf{t}_1\}\{P_2|\mathbf{t}_2\} = \{P_1 P_2|\mathbf{t}_1 + P_1\mathbf{t}_2\}. \tag{S2}$$

Note that here the point group element changes the translation element $\mathbf{t}_2$, which is different from the usual direct product $g_1 g_2 = \{P_1 P_2|\mathbf{t}_1 + \mathbf{t}_2\}$.

### B. Tight-binding basis

The space group, Wyckoff position and orbitals on the sites determine the symmetry representations in the Brillouin zone, which is important for defining the basis for the tight-binding representation. We construct the band representations following Ref. [1].

To construct the tight-binding basis, we start by considering one lattice site (sublattice) at position $\mathbf{q}$. We can now define the site-symmetry group, $G_{\mathbf{q}}$, which includes all space group operations $g$ that leave the site invariant, $G_{\mathbf{q}} = \{g|g\mathbf{q} = \mathbf{q}\}$. Note that the site-symmetry group is a subgroup (or isomorphic to a subgroup) of the full space group. The orbital at site $\mathbf{q}$ belong to an irreducible representation (IR), $\rho$, of the site symmetry group. The elements of the space group $g$ that are not in the site symmetry group $G_{\mathbf{q}}$ move the site $\mathbf{q}$ to an orbit of other sites (defined by the Wyckoff position).

We label our starting site (sublattice) at $\mathbf{R} = \mathbf{0}$, site 1 and orbital $\mu$ as $|\phi_{\mathbf{0},1,\mu}\rangle$. Therefore, to obtain the complete basis we have to define $|\phi_{\mathbf{R},\alpha,\mu}\rangle$, where $\mathbf{R}, \alpha, \mu$ denote the unit cell, sublattice and orbital, respectively. With this purpose, we need to introduce a group theory tool, the coset decomposition. Given a group $G$ and a subgroup $H$, then the group $G$ can always be partitioned as

$$G = \bigcup_{\alpha} g_\alpha H = g_1 H + g_2 H + \cdots + g_n H, \tag{S3}$$

with $g_\alpha$ denoting the coset representatives. We can illustrate this concept by considering the point group $D_{4h}$ and the subgroup $D_{2h}$. In this example, the coset decomposition is given by $G = ED_{2h} + C_4D_{2h}$, where the coset representatives $E$ and $C_4$ denote the identity and the four-fold rotation around the $z$ axis, respectively. In our case, $G$ is the full space group and the subgroup $H = G_{\mathbf{q}}$ is the site symmetry group. Thus, the coset decomposition corresponds to

$$G = \bigcup_\alpha g_\alpha (G_{\mathbf{q}} \ltimes \mathbb{Z}^n), \tag{S4}$$

where $\ltimes$ corresponds to the semi-direct product and $\mathbb{Z}^n$ is the translation group ($n$ is the dimension of the space). Hence, the tight-binding basis is defined as

$$|\phi_{\mathbf{R},\alpha,\mu}\rangle \equiv \{E|\mathbf{R}\} g_\alpha |\phi_{\mathbf{0},1,\mu}\rangle . \tag{S5}$$

### C. Symmetries acting on tight-binding basis in real space

Having defined the basis, the next step to derive the form for the tight-binding model is to examine how symmetries act on our basis. We begin the analysis in real space, and in the next section we will focus on the symmetries in momentum space.

Recalling that the orbital belongs to the IR $\rho$ of the site-symmetry group, the elements $g$ in the site-symmetry group do not move the site, only transform the orbitals, and therefore satisfy

$$\langle \phi_{\mathbf{0},1,\nu}| g |\phi_{\mathbf{0},1,\mu}\rangle = [\rho(g)]_{\nu\mu}, \quad \text{for} \quad g \in G_{\mathbf{q}}. \tag{S6}$$

If we now consider the elements $h = \{P|\mathbf{v}\}$ which are not in the site-symmetry group and the tight-binding basis in Eq. (S5),

$$h |\phi_{\mathbf{R},\alpha,\mu}\rangle = \{P|\mathbf{v}\}\{E|\mathbf{R}\} g_\alpha |\phi_{\mathbf{0},1,\mu}\rangle = \{P|\mathbf{v} + P\mathbf{R}\} g_\alpha |\phi_{\mathbf{0},1,\mu}\rangle = \{E|P\mathbf{R}\}\{P|\mathbf{v}\} g_\alpha |\phi_{\mathbf{0},1,\mu}\rangle , \tag{S7}$$

where we have used the semi-direct product of space group elements introduced in Eq. (S2). To identify how the elements $h$ transform the coset $g_\alpha$, we can now use the coset decomposition: $hg_\alpha = \{E|\mathbf{t}\} g_\beta g$, where $\mathbf{t}$ is a lattice vector, $g_\beta$ is the coset representative for the site $\beta$, and $g \in G_{\mathbf{q}}$ corresponds to an element of the site-symmetry group transforming the orbitals. Hence, we obtain

$$\begin{aligned} h |\phi_{\mathbf{R},\alpha,\mu}\rangle &= \{E|P\mathbf{R}\} h g_\alpha |\phi_{\mathbf{0},1,\mu}\rangle = \{E|P\mathbf{R}\}\{E|\mathbf{t}\} g_\beta g |\phi_{\mathbf{0},1,\mu}\rangle = \{E|P\mathbf{R}\}\{E|\mathbf{t}\} g_\beta |\phi_{\mathbf{0},1,\nu}\rangle \langle \phi_{\mathbf{0},1,\nu}| g |\phi_{\mathbf{0},1,\mu}\rangle = \\ &= \{E|P\mathbf{R} + \mathbf{t}\} g_\beta |\phi_{\mathbf{0},1,\nu}\rangle [\rho(g)]_{\nu\mu} = |\phi_{P\mathbf{R}+\mathbf{t},\beta,\nu}\rangle [\rho(g)]_{\nu\mu}, \end{aligned} \tag{S8}$$

where we have used Eq. (S6) for the site-symmetry elements. Here, $\mathbf{t}$, $\beta$ and $g$ are defined by the coset decomposition. Defining $\mathbf{R}' = P\mathbf{R} + \mathbf{t}$ and introducing the transformation matrix $[U_h]_{\beta\nu,\alpha\mu}$ with the sublattice degrees of freedom, the previous expression can generally be written as

$$h |\phi_{\mathbf{R},\alpha,\mu}\rangle = [U_h]_{\beta\nu,\alpha\mu} |\phi_{\mathbf{R}',\beta,\nu}\rangle . \tag{S9}$$

This shows that the transformation matrix $[U_h]_{\beta\nu,\alpha\mu}$ is independent of $\mathbf{R}$ and $\mathbf{R}'$, which will simplify the transformation of the tight-binding basis in momentum space, as we shown in the following subsection. The previous equation also reveals that the key step to determine the tight-binding model is to identify the form of the matrix $U$, which we derive for the Lieb lattice model in Sec. S2.

In the basis introduced in Eq. (S5), the tight-binding Hamiltonian corresponds to

$$H_{\alpha\mu,\beta\nu}(\mathbf{R}) \equiv \langle \phi_{\mathbf{R}',\alpha,\mu}| H |\phi_{\mathbf{R}'+\mathbf{R},\beta,\nu}\rangle = \langle \phi_{0,\alpha,\mu}| H |\phi_{\mathbf{R},\beta,\nu}\rangle . \tag{S10}$$

Therefore, due to translational invariance, the Hamiltonian is fully specified by the unit cell, the sublattice and the orbital indices.

### D. Symmetries acting on tight-binding basis in momentum space

There are two relevant choices for the Fourier transform to define the tight-binding basis in momentum space $|\phi_{\mathbf{k},\alpha,j}\rangle$. In the first choice,

$$|\tilde{\varphi}_{\mathbf{k},\alpha,\nu}\rangle = \sum_{\mathbf{k}} e^{i\mathbf{k}\cdot\mathbf{R}} |\phi_{\mathbf{R},\alpha,\nu}\rangle . \tag{S11}$$

Therefore, the Hamiltonian is periodic in the Brillouin zone:

$$\tilde{H}_{\alpha\mu,\beta\nu}(\mathbf{k}) \equiv \langle\tilde{\varphi}_{\mathbf{k},\alpha,\mu}|\, H\, |\tilde{\varphi}_{\mathbf{k},\beta,\nu}\rangle = \sum_{\mathbf{R}} e^{i\mathbf{k}\cdot\mathbf{R}} H_{\alpha\mu,\beta\nu}(\mathbf{R}). \tag{S12}$$

In the second choice,

$$|\varphi_{\mathbf{k},\alpha,\nu}\rangle = \sum_{\mathbf{k}} e^{i\mathbf{k}\cdot(\mathbf{R}+\mathbf{r}_\alpha)}\, |\phi_{\mathbf{R},\alpha,\nu}\rangle\,, \tag{S13}$$

and therefore the sublattice position $\mathbf{r}_\alpha$ enters in the transformation. The Hamiltonian is then given by

$$H_{\alpha\mu,\beta\nu}(\mathbf{k}) \equiv \langle\varphi_{\mathbf{k},\alpha,\mu}|\, H\, |\varphi_{\mathbf{k},\beta,\nu}\rangle = \sum_{\mathbf{R}} e^{i\mathbf{k}\cdot(\mathbf{R}-\mathbf{r}_\alpha+r_\beta)} H_{\alpha\mu,\beta\nu}(\mathbf{R}). \tag{S14}$$

Note that now the Hamiltonian is not periodic in the Brillouin zone,

$$H_{\alpha\mu,\beta\nu}(\mathbf{k}+\mathbf{G}) = \sum_{\mathbf{R}} e^{i\mathbf{G}\cdot(\mathbf{r}_\beta-\mathbf{r}_\alpha)} H_{\alpha\mu,\beta\nu}(\mathbf{k}). \tag{S15}$$

Here, we use this choice, which is more convenient since the symmetries without translations are momentum independent, as we will now show.

First, we have to determine how the space group symmetries $g$ act on the Fourier-transformed tight-binding basis. To this end, we use the result in Eq. (S9),

$$g\,|\varphi_{\mathbf{k},\alpha,\mu}\rangle = \sum_{\mathbf{R}} e^{i\mathbf{k}\cdot(\mathbf{R}+\mathbf{r}_\alpha)} g\,|\phi_{\mathbf{R},\alpha,\mu}\rangle = \sum_{\mathbf{R}} e^{i(P\mathbf{k})\cdot(P(\mathbf{R}+\mathbf{r}_\alpha))}\,|\phi_{\mathbf{R}',\beta,\nu}\rangle\,[U_g]_{\beta\nu,\alpha\mu}, \tag{S16}$$

where we have also used that $e^{i\mathbf{k}\cdot(\mathbf{R}+\mathbf{r}_\alpha)} = e^{i\mathbf{k}\cdot(P^{-1}P(\mathbf{R}+\mathbf{r}_\alpha))} = e^{i(P\mathbf{k})\cdot(P(\mathbf{R}+\mathbf{r}_\alpha))}$. Recalling the definition for $\mathbf{R}'$ introduced below Eq. (S9), we can rewrite $P(\mathbf{R}+\mathbf{r}_\alpha) = \mathbf{R}'+\mathbf{r}_\beta-\mathbf{t}$. Thus, we obtain

$$g\,|\varphi_{\mathbf{k},\alpha,\mu}\rangle = |\varphi_{P\mathbf{k},\beta,\mu}\rangle\, e^{-i(P\mathbf{k})\cdot\mathbf{t}}[U_g]_{\beta\nu,\alpha\mu}. \tag{S17}$$

Notably, the translations in momentum space become a phase, and the matrix representation is then independent on the phase and momentum.

We now demonstrate that the previous momentum-independent matrix representations determine the symmetries of the tight-binding model. To this end, we first show how the tight-binding Hamiltonian transforms under the space group symmetries $g$:

$$H_{\alpha\mu,\beta\nu}(\mathbf{k}) \equiv \langle\varphi_{\mathbf{k},\alpha,\mu}|\, H\, |\varphi_{\mathbf{k},\beta,\nu}\rangle = \langle\varphi_{\mathbf{k},\alpha,\mu}|\, gHg^{-1}\, |\varphi_{\mathbf{k},\beta,\nu}\rangle\,. \tag{S18}$$

Introducing the identity in our basis, we can then use Eq. (S17),

$$H_{\alpha\mu,\beta\nu}(\mathbf{k}) = \sum_{\substack{\mathbf{k}',\mathbf{k}''\\ \alpha',\beta',\mu',\nu'}} \langle\varphi_{\mathbf{k},\alpha,\mu}|\, g\, |\varphi_{\mathbf{k}',\alpha',\mu'}\rangle\, \langle\varphi_{\mathbf{k}',\alpha',\mu'}|\, H\, |\varphi_{\mathbf{k}'',\beta',\nu'}\rangle\, \langle\varphi_{\mathbf{k}'',\beta',\nu'}|\, g^{-1}\, |\varphi_{\mathbf{k},\beta,\nu}\rangle \tag{S19}$$

$$= \sum_{\substack{\mathbf{k}',\mathbf{k}''\\ \alpha',\beta',\mu',\nu'}} \left(\delta_{P\mathbf{k}',\mathbf{k}} e^{-iP\mathbf{k}'\cdot\mathbf{t}}[U_g]_{\alpha\mu,\alpha'\mu'}\right)\left(\delta_{\mathbf{k}',\mathbf{k}''} H_{\alpha'\mu',\beta'\nu'}(\mathbf{k})\right)\left(\delta_{P\mathbf{k}'',\mathbf{k}}[U_g^{-1}]_{\beta'\nu',\beta\nu} e^{iP\mathbf{k}''\cdot\mathbf{t}}\right) \tag{S20}$$

$$= \sum_{\alpha',\beta',\mu',\nu'} [U_g]_{\alpha\mu,\alpha'\mu'} H_{\alpha'\mu',\beta'\nu'}(P^{-1}\mathbf{k})[U_g^{-1}]_{\beta'\nu',\beta\nu}. \tag{S21}$$

Importantly, from this result we can conclude that the momentum independent matrices $U_g$ determine how the tight-binding Hamiltonian transform under symmetries:

$$H_{\alpha\mu,\beta\nu}(\mathbf{k}) = [U_g H(P^{-1}\mathbf{k})U_g^{-1}]_{\alpha\mu,\beta\nu}. \tag{S22}$$

### E. Tight-binding Hamiltonian rules

Therefore, we have derived the two key tight-binding Hamiltonian rules:

1. Periodicity of the Hamiltonian:

$$H_{\alpha\mu,\beta\nu}(\mathbf{k}+\mathbf{G}) = \sum_{\mathbf{R}} e^{i\mathbf{G}\cdot(\mathbf{r}_\beta-\mathbf{r}_\alpha)} H_{\alpha\mu,\beta\nu}(\mathbf{k}). \tag{S23}$$

2. Transformation of the Hamiltonian under space group symmetries:

$$H(\mathbf{k}) = U_g H(P^{-1}\mathbf{k}) U_g^{-1}. \tag{S24}$$

## S2. EXAMPLE: INVERSE LIEB LATTICE MODEL

In this section, we follow the procedure presented above to derive the tight-binding Hamiltonian for the inverse Lieb lattice model. We start by introducing the space group and the generators and defining the tight-binding basis. Then, we identify the matrices describing the space group symmetries and use the tight-binding rules to derive the final form for the tight-binding model.

### A. Space group and generators

The space group for the inverse Lieb lattice is $P_{4/mmm}$, which corresponds to space group number 123. The position of the magnetic atoms is given by $(0, 1/2)$ and $(1/2, 0)$, which is labeled by the Wyckoff position 2f in the Bilbao Crystallographic Server [2, 3]. The generators of this space group are $\{E|\mathbf{t}_1\}$, $\{E|\mathbf{t}_2\}$, $\{C_{4z}|\mathbf{0}\}$, $\{I|\mathbf{0}\}$, $\{m_x|\mathbf{0}\}$ and $\{m_z|\mathbf{0}\}$, with $E$, $C_{4z}$, $I$, $m_x$ and $m_z$ denoting the identity, four-fold rotation, inversion and mirrors across the $x$ and $z$ axis, respectively. Equivalently, the space group can be written as $P_{4/mmm} = D_{4h} \ltimes Z^2$, where $Z^2$ denotes the translation group generated by $\{E|\mathbf{t}_1\}$ and $\{E|\mathbf{t}_2\}$, with $\mathbf{t}_1 = (1, 0)$ and $\mathbf{t}_2 = (0, 1)$ for the square lattice.

For the orbital degree of freedom, motivated by the $AV_2Q_2O$-family altermagnets, we construct the following two tight-binding models: 1) $xy$ orbitals on the Wyckoff position and 2) $xz$ orbitals on one sublattice and $yz$ on the other.

### B. Tight-binding basis

In our example, the site-symmetry group corresponds to $P_{mmm} = D_{2h} \ltimes Z^2$, which is a subgroup of the full group $P_{4/mmm} = D_{4h} \ltimes Z^2$. The elements of the full space group $P_{4/mmm}$ that are not in the site-symmetry group $P_{mmm}$ move the site to the other site defined by the Wyckoff position 2f.

In the case of two sublattices, there are only two coset representatives. Since the two sublattices are related by $\{C_{4z}|\mathbf{0}\}$, following the coset decomposition introduced in Eq. (S3), the space group can be written as

$$P_{4/mmm} = \{E|\mathbf{0}\}P_{mmm} + \{C_{4z}|\mathbf{0}\}P_{mmm}. \tag{S25}$$

In this case, the coset representatives $g_1 = \{E|\mathbf{0}\}$ and $g_2 = \{C_{4z}|\mathbf{0}\}$ define the basis in Eq. (S5). With this purpose, we start with the lattice site at $(1/2, 0)$ and label it as $|\phi_{\mathbf{0},1,\mu}\rangle$. The site-symmetry group $G_{\mathbf{q}}$ is isomorphic to $D_{2h}$, which contains the following point group operations: $\{E, I, m_x, m_y, m_z, C_{2x}, C_{2y}, C_{2z}\}$. The orbitals $\mu$ belong to an IR $\rho^\mu$ of $G_{\mathbf{q}}$. Specifically, $xy$ orbitals belong to $\rho^{xy} = B_{1g}$ IR, while $xz$ ($yz$) orbitals belong to $\rho^{xz} = B_{2g}$ ($\rho^{yz} = B_{3g}$) IR. Therefore, since $\{C_4|\mathbf{0}\}$ generates the other site in the Wyckoff position, the tight-binding basis is defined as

$$|\phi_{\mathbf{0},2,\mu}\rangle \equiv \{C_4|\mathbf{0}\}\, |\phi_{\mathbf{0},1,\mu}\rangle\,. \tag{S26}$$

### C. Matrix representation for the space group symmetries

Once we have defined the tight-binding basis, the goal now is to derive the symmetry of the Pauli matrices $\tau_i^\mu$ in the sublattice basis for the orbital $\mu$. To this end, we need to obtain the matrix form of the space group generators in the sublattice basis $\{|\phi_{\mathbf{R},1,\mu}\rangle\,, |\phi_{\mathbf{R},2,\mu}\rangle\}$.

First, we consider the four-fold rotation $\{C_4|\mathbf{0}\}$, and apply this symmetry to our basis elements, recalling that we only know how the basis transforms under site-symmetry operations. Therefore, we obtain

$$\{C_4|\mathbf{0}\}\,|\phi_{\mathbf{R},1,\mu}\rangle \equiv |\phi_{\mathbf{R},2,\mu}\rangle\,, \tag{S27}$$

$$\{C_4|\mathbf{0}\}\,|\phi_{\mathbf{R},2,\mu}\rangle = \{C_4|\mathbf{0}\}^2\,|\phi_{\mathbf{R},1,\mu}\rangle = \{C_{2z}|\mathbf{0}\}\,|\phi_{\mathbf{R},1,\mu}\rangle = \{E|-10\}\{C_{2z}|10\}\,|\phi_{\mathbf{R},1,\mu}\rangle = \{E|-10\}\rho^{\mu}(C_{2z})\,|\phi_{\mathbf{R},1,\mu}\rangle\,. \tag{S28}$$

Here, we have used that $\{C_{2z}|10\}$ is a site-symmetry operation, while $\{C_{2z}|\mathbf{0}\}$ is not. Hence, we can use the following rule for the commutation of space group elements to obtain the previous result: $\{P|\mathbf{v}\}\{E|\mathbf{R}\} = \{E|P\mathbf{R}\}\{P|\mathbf{v}\}$. For our set of orbitals, $\rho^{xy}(C_{2z}) = +1$ while $\rho^{xz/yz}(C_{2z}) = -1$. As a consequence, the matrix representation for the four-fold rotation depends on the orbital considered, and corresponds to

$$U^{\mu}_{\{C_4|\mathbf{0}\}} = \begin{pmatrix} 0 & 1 \\ \rho^{\mu}(C_{2z}) & 0 \end{pmatrix} = \begin{cases} \tau_x & \text{for } xy \text{ orbitals,} \\ i\tau_y & \text{for } xz/yz \text{ orbitals.} \end{cases} \tag{S29}$$

The other space group generators of $P_{4/mmm}$ do not exchange the two sublattices, and therefore the matrix representation is diagonal. We now analyze in detail the $\{m_x|\mathbf{0}\}$ symmetry operation. In this case, recalling that the basis is defined by Eq. (S26), the symmetry applied to our basis elements corresponds to

$$\{m_x|\mathbf{0}\}\,|\phi_{\mathbf{R},1,\mu}\rangle = \rho^{\mu}(m_x)\,|\phi_{\mathbf{R},1,\mu}\rangle\,, \tag{S30}$$

$$\{m_x|\mathbf{0}\}\,|\phi_{\mathbf{R},2,\mu}\rangle = \{m_x|\mathbf{0}\}\{C_4|\mathbf{0}\}\,|\phi_{\mathbf{R},1,\mu}\rangle = \{C_4|\mathbf{0}\}\{m_y|\mathbf{0}\}\,|\phi_{\mathbf{R},1,\mu}\rangle = \rho^{\mu}(m_y)\{C_4|\mathbf{0}\}\,|\phi_{\mathbf{R},1,\mu}\rangle = \rho^{\mu}(m_y)\,|\phi_{\mathbf{R},2,\mu}\rangle\,. \tag{S31}$$

Note that $m_x$ and $C_4$ do not commute, and we obtain the mirror $m_y$ instead. Therefore, this gives the following matrix representation:

$$U^{\mu}_{\{m_x|\mathbf{0}\}} = \begin{pmatrix} \rho^{\mu}(m_x) & 0 \\ 0 & \rho^{\mu}(m_y) \end{pmatrix} = \frac{1}{2}[\rho^{\mu}(m_x)+\rho^{\mu}(m_y)]\tau_0+\frac{1}{2}[\rho^{\mu}(m_x)-\rho^{\mu}(m_y)]\tau_z = \begin{cases} \tau_0 & \text{for } xy \text{ orbitals,} \\ \tau_z & \text{for } xz/yz \text{ orbitals.} \end{cases} \tag{S32}$$

Similarly, for the generators $\{I|\mathbf{0}\}$ and $\{m_z|\mathbf{0}\}$, the matrix representation corresponds to

$$U^{\mu}_{\{I|\mathbf{0}\}} = \begin{pmatrix} \rho^{\mu}(I) & 0 \\ 0 & \rho^{\mu}(I) \end{pmatrix}, \quad U^{\mu}_{\{m_z|\mathbf{0}\}} = \begin{pmatrix} \rho^{\mu}(m_z) & 0 \\ 0 & \rho^{\mu}(m_z) \end{pmatrix}. \tag{S33}$$

In these cases, the representation is proportional to $\tau_0$ for the $xy$ and $xz/yz$ orbitals and may differ only by an overall minus sign.

Hence, we can now identify the symmetry of the operators $\tau_i^{\mu}$ in the tight-binding Hamiltonian by calculating $U_g^{\mu}\tau_i^{\mu}U_g^{\mu-1}$. The result is shown in Tables S1-S2 for $xy$ and $xz/yz$ orbitals. The key result is that the symmetry of the $\tau_x$ and $\tau_y$ operators is different for the two sets of orbitals, which implies that the tight-binding Hamiltonian will have a different form. Another important result is that the symmetry of $\tau_z$ does not depend on the orbitals considered, and therefore the symmetry of the altermagnetic spin splitting is $B_{1g}$ in both cases.

Table S1. Irreducible representations of the operators in sublattice space for $xy$ orbitals.

| $xy$ | $E$ | $C_4$ | $m_x$ | $I$ | $m_z$ | IR |
|---|---|---|---|---|---|---|
| $\tau_0$ | 1 | 1 | 1 | 1 | 1 | $A_{1g}$ |
| $\tau_x$ | 1 | 1 | 1 | 1 | 1 | $A_{1g}$ |
| $\tau_y$ | 1 | -1 | 1 | 1 | 1 | $B_{1g}$ |
| $\tau_z$ | 1 | -1 | 1 | 1 | 1 | $B_{1g}$ |

Table S2. Irreducible representations of the operators in sublattice space for $xz/yz$ orbitals.

| $xz/yz$ | $E$ | $C_4$ | $m_x$ | $I$ | $m_z$ | IR |
|---|---|---|---|---|---|---|
| $\tau_0$ | 1 | 1 | 1 | 1 | 1 | $A_{1g}$ |
| $\tau_x$ | 1 | -1 | -1 | 1 | 1 | $B_{2g}$ |
| $\tau_y$ | 1 | 1 | -1 | 1 | 1 | $A_{2g}$ |
| $\tau_z$ | 1 | -1 | 1 | 1 | 1 | $B_{1g}$ |

### D. Tight-binding Hamiltonian

Once we have determined the symmetry of the $\tau_i$ operators, we can finally obtain the tight-binding Hamiltonian. The general form for the microscopic model is given by

$$H_0^{\mu}(\mathbf{k}) = \sum_{j,l} f_{jl}(\mathbf{k})\tau_j^{\mu}\sigma_l, \tag{S34}$$

where $\tau_j^{\mu}$ corresponds to the sublattice space matrices for the orbital $\mu$ and $\sigma_l$ denotes the spin degree of freedom.

Let us focus first on the model for $xy$ orbitals. To obtain the right form for the momentum-dependent factors $f_{jl}(\mathbf{k})$ entering in Eq. (S34), we use that the Hamiltonian must transform trivially under all space group operations. Therefore, since inversion symmetry is present and $\tau_y$ and the spin matrices $\sigma_l$ break time-reversal symmetry, they correspond to the spin-orbit coupling terms. In addition, we see that for $\tau_z$ the momentum-dependent prefactor must transform like $\cos k_x - \cos k_y \sim B_{1g}$. Hence, the tight-binding Hamiltonian for $xy$ orbitals can be written as

$$H_0^{xy}(\mathbf{k}) = \varepsilon_{0,\mathbf{k}}\tau_0 + t_x^{xy}\cos\frac{k_x}{2}\cos\frac{k_y}{2}\tau_x^{xy} + t_{z,1}^{xy}(\cos k_x - \cos k_y)\tau_z + \lambda_z^{xy}\sin\frac{k_x}{2}\sin\frac{k_y}{2}\tau_y^{xy}\sigma_z, \tag{S35}$$

where $\varepsilon_{0,\mathbf{k}} = t_1(\cos k_x + \cos k_y) + t_2\cos k_x\cos k_y - \mu$. Here, we have used that in order to satisfy the periodicity of the Hamiltonian in Eq. (S23), the hoppings between opposite sublattices (described by the off-diagonal matrices $\tau_x$ and $\tau_y$) must have an additional momentum dependence corresponding to the factors of 1/2. Additionally, using the transformation properties in Table S1, we see that the previous form satisfies Eq. (S24). Note that the SOC terms $\sigma_x$ and $\sigma_y$ do not appear since we are considering a two-dimensional model. Since $\sigma_z \sim A_{2g}$ in $D_{4h}$ and $\tau_y^{xy} \sim B_{1g}$, the momentum dependence must transform as $A_{2g}\otimes B_{1g} = B_{2g} \sim k_x k_y$.

Similarly, we now focus on the model for $xz/yz$ sublattices. In this case, as shown in Table S2, the symmetries for the $\tau_x$ and $\tau_y$ operators are different. Therefore, following similar arguments as in the previous case, the tight-binding Hamiltonian corresponds to

$$H_0^{xz/yz}(\mathbf{k}) = \varepsilon_{0,\mathbf{k}}\tau_0 + t_x^{xz/yz}\sin\frac{k_x}{2}\sin\frac{k_y}{2}\tau_x^{xz/yz} + t_{z,1}^{xz/yz}(\cos k_x - \cos k_y)\tau_z + \lambda_z^{xz/yz}\cos\frac{k_x}{2}\cos\frac{k_y}{2}\tau_y^{xz/yz}\sigma_z. \tag{S36}$$

Thus, we have derived the tight-binding Hamiltonians for $xy$ and $xz/yz$ orbitals that we present in Eqs. (1)-(5) of the main text.

## S3. SPLITTING OF SINGLE-PARTICLE HAMILTONIANS FOR ORBITALS BELONGING TO 2D IRREDUCIBLE REPRESENTATIONS

Motivated by the $AV_2Q_2O$ family of materials, in the main text we consider a model with $xz/yz$ orbitals. In the tetragonal point group $D_{4h}$, these orbitals belong to the 2D $E_g$ IR. However, since the site symmetry group is orthorhombic ($D_{2h}$), which does not contain 2D IRs, the $4\times 4$ tight-binding model splits into two blocks. As a consequence, we can consider a model with only $xz$ orbitals on site 1 and $yz$ orbitals on site 2, since the other $xz/yz$ bands are far away from the Fermi energy.

Here, we analyze this problem more generally and consider orbitals belonging to the 2D IRs of space groups that allow altermagnetism with two sites, as identified in Ref. [4]. Notably, from the previous space groups, only tetragonal, rhombohedral, hexagonal and cubic space groups contain 2D IR. In the following, we analyze all these cases in detail for the different point groups ($P$) and site-symmetry groups ($S$). We address two questions: whether the orbitals belonging to 2D IRs split into two blocks, and whether the theories for 1D orbitals in the site-symmetry group can take different forms.

1) Tetragonal space group, $P = D_{4h}$, $S = D_{2h}$: The symmetry relating the two sites is the four-fold rotation $C_{4z}$. The key symmetries are the mirrors $m_x$ and $m_y$, which satisfy $m_y = C_{4z}^{-1}m_x C_{4z}$. Therefore, this gives rise to two possibilities for the symmetries of $(\tau_x, \tau_y, \tau_z)$: a) $(A_{1g}, B_{1g}, A_{1g}\otimes B_{1g} = B_{1g})$ or b) $(B_{2g}, A_{2g}, B_{2g}\otimes A_{2g} = B_{1g})$. Here, we have used that the symmetry of $\tau_z$ is fixed by the product of the IR for $\tau_x$ and $\tau_y$. As a consequence, there are two possible theories for the orbitals in $D_{2h}$, depending on whether the orbitals are even or odd under the mirror symmetry $m_x$. Our example for the $AV_2Q_2O$ family of materials belongs to this class. In this case, the $xy$ ($xz/yz$) orbitals are even (odd) under $m_x$, and therefore they belong to different theories.

2) Tetragonal space group, $P = C_{4h}$, $S = C_{2h}$: The symmetry relating the two sites is also $C_{4z}$. The site-symmetry group does not contain 2D IRs, and therefore the tight-binding Hamiltonian for orbitals belonging to 2D IRs (such as the $xz/yz$) splits into two blocks. However, the point group $C_{2h}$ only contains two 1D IR and the relevant mirror $m_z$ satisfies $m_z = C_{4z}^{-1}m_z C_{4z}$. As a consequence, the tight-binding model for all 1D orbitals has the same form.

3) Tetragonal space group, $P = D_{4h}$, $S = C_{4h}$: The two sites are related by the mirror symmetries $m_x$ and $m_{xy}$ (diagonal mirror). The orbitals belonging to 2D IRs in $P$ are not split in $S$. In this case, $m_z = m_x^{-1} m_z m_x$, and therefore the tight-binding model for all 1D orbitals has the same form.

4) Rhombohedral space group, $P = D_{3d}$, $S = S_6$: The two sites are related by the diagonal mirror $m_{xy}$. Similar to the previous case, there is no splitting of orbitals in $S$ belonging to 2D IRs in $P$. Since there is only one 1D IR in $S_6$, all 1D orbitals have the same tight-binding model.

5) Hexagonal space group, $P = C_{6h}$, $S = S_6$: The two sites are related by the two-fold rotation $C_{2z}$. Similar to the previous case, there is no splitting of the orbitals belonging to 2D IRs and all 1D orbitals have the same tight-binding model.

6) Hexagonal space group, $P = D_{6h}$, $S = C_{6h}$: The two sites are related by the mirror symmetries $m_x$ and $m_{xy}$. There is no splitting of orbitals belonging to 2D IRs. Since the six-fold symmetry $C_{6z}$ satisfies $C_{6z} = m_x^{-1} C_{6z} m_x$, all 1D orbitals have the same tight-binding form.

7) Hexagonal space group, $P = D_{6h}$, $S = D_{3d}$: The symmetry relating the two sites is the six-fold rotation $C_{6z}$. In this case, for the diagonal symmetry we obtain $m_{xy} = C_{6z}^{-1} m_{xy} C_{6z}$, and therefore all the tight-binding models for 1D orbitals have the same form.

8) Cubic space group, $P = O_h$, $S = D_{3d}$: The four-fold rotation $C_{4z}$ relates the two sublattices. Similar to the previous case, $m_{xy} = C_{4z}^{-1} m_{xy} C_{4z}$, and all the tight-binding models for 1D orbitals have the same form.

## S4. ANALYTIC EXPRESSIONS FOR THE FERROMAGNETIC AND ALTERMAGNETIC SUSCEPTIBILITIES

To gain insight into when the altermagnetic instability is favored over the ferromagnetic instability, we have derived analytic expressions for the ferromagnetic and altermagnetic susceptibilities. To this end, we consider the normal-state Hamiltonian

$$H_0 = \varepsilon_{0,\mathbf{k}} \tau_0 + t_{x,\mathbf{k}} \tau_x + t_{z,\mathbf{k}} \tau_z, \tag{S37}$$

with eigenvalues $E_{\mathbf{k}}^{\pm} = \varepsilon_{0,\mathbf{k}} \pm \sqrt{t_{x,\mathbf{k}}^2 + t_{z,\mathbf{k}}^2}$. The ferromagnetic susceptibility, which diverges at $\mathbf{q} \to 0$ close to a ferromagnetic instability, is given by

$$\chi_0^{\mathrm{FM}}(\mathbf{q}, iq_n) = -\int_0^\beta e^{iq_n\tau} \langle T_\tau S_{\mathbf{q}}(\tau) S_{-\mathbf{q}}(0) \rangle, \tag{S38}$$

where $S_{\mathbf{q}} = \frac{1}{N} \sum_{\mathbf{k}} \Psi_{\mathbf{k}+\mathbf{q}}^\dagger \tau_0 \Psi_{\mathbf{k}}$, with the spinor $\Psi_{\mathbf{k}} = (\psi_{\mathbf{k},1} \; \psi_{\mathbf{k},2})^T$ in the sublattice basis. In the band basis, it can be rewritten as

$$\chi_0^{\mathrm{FM}}(\mathbf{q}, iq_n) = -\frac{1}{N} \sum_{\mathbf{k},a,b} \left| \langle u_{\mathbf{k}}^a | u_{\mathbf{k}+\mathbf{q}}^b \rangle \right|^2 \frac{f(E_{\mathbf{k}}^a) - f(E_{\mathbf{k}+\mathbf{q}}^b)}{iq_n + E_{\mathbf{k}}^a - E_{\mathbf{k}+\mathbf{q}}^b}. \tag{S39}$$

We can use the analytic form for the eigenvectors of the transformation from the sublattice to the band basis, which correspond to

$$|u_{\mathbf{k}}^+\rangle = \begin{pmatrix} \cos\frac{\theta_{\mathbf{k}}}{2} \\ -\sin\frac{\theta_{\mathbf{k}}}{2} \end{pmatrix}, \quad |u_{\mathbf{k}}^-\rangle = \begin{pmatrix} \sin\frac{\theta_{\mathbf{k}}}{2} \\ \cos\frac{\theta_{\mathbf{k}}}{2} \end{pmatrix}, \tag{S40}$$

with $\cos\theta_{\mathbf{k}} = \frac{t_{z,\mathbf{k}}}{\sqrt{t_{x,\mathbf{k}}^2 + t_{z,\mathbf{k}}^2}}$ and $\sin\theta_{\mathbf{k}} = \frac{t_{x,\mathbf{k}}}{\sqrt{t_{x,\mathbf{k}}^2 + t_{z,\mathbf{k}}^2}}$. Therefore, the ferromagnetic susceptibility can be written as

$$\begin{aligned} \chi_0^{\mathrm{FM}}(\mathbf{q}, iq_n) &= -\left( \cos\frac{\theta_{\mathbf{k}} - \theta_{\mathbf{k}+\mathbf{q}}}{2} \right)^2 (f_{++} + f_{--}) - \left( \sin\frac{\theta_{\mathbf{k}} - \theta_{\mathbf{k}+\mathbf{q}}}{2} \right)^2 (f_{+-} + f_{-+}) \\ &= -\frac{1 + \cos(\theta_{\mathbf{k}} - \theta_{\mathbf{k}+\mathbf{q}})}{2} (f_{++} + f_{--}) - \frac{1 - \cos(\theta_{\mathbf{k}} - \theta_{\mathbf{k}+\mathbf{q}})}{2} (f_{+-} + f_{-+}), \end{aligned} \tag{S41}$$

where we have defined $f_{ab} = \frac{f(E_{\mathbf{k}}^a) - f(E_{\mathbf{k}+\mathbf{q}}^b)}{iq_n + E_{\mathbf{k}}^a - E_{\mathbf{k}+\mathbf{q}}^b}$.

The altermagnetic susceptibility corresponds to [4]

$$\chi_0^{\mathrm{AM}}(\mathbf{q}, iq_n) = -\int_0^\beta e^{iq_n\tau} \langle T_\tau \tilde{S}_{\mathbf{q}}(\tau) \tilde{S}_{-\mathbf{q}}(0) \rangle, \tag{S42}$$

where now $\tilde{S}_{\mathbf{q}} = \frac{1}{N}\sum_{\mathbf{k}} \Psi^{\dagger}_{\mathbf{k}+\mathbf{q}}\tau_z\Psi_{\mathbf{k}}$. In the band basis,

$$\chi_0^{\mathrm{AM}}(\mathbf{q}, iq_n) = -\frac{1}{N}\sum_{\mathbf{k},a,b} \left|\langle u^a_{\mathbf{k}}|\,\tau_z\,|u^b_{\mathbf{k}+\mathbf{q}}\rangle\right|^2 \frac{f(E^a_{\mathbf{k}})-f(E^b_{\mathbf{k}+\mathbf{q}})}{iq_n+E^a_{\mathbf{k}}-E^b_{\mathbf{k}+\mathbf{q}}}. \tag{S43}$$

Using the form for the eigenvector in Eq. (S40), the altermagnetic susceptibility can generally be written as

$$\begin{aligned}\chi_0^{\mathrm{AM}}(\mathbf{q}, iq_n) &= -\left(\cos\frac{\theta_{\mathbf{k}}+\theta_{\mathbf{k}+\mathbf{q}}}{2}\right)^2(f_{++}+f_{--}) - \left(\sin\frac{\theta_{\mathbf{k}}+\theta_{\mathbf{k}+\mathbf{q}}}{2}\right)^2(f_{+-}+f_{-+})\\ &= -\frac{1+\cos(\theta_{\mathbf{k}}+\theta_{\mathbf{k}+\mathbf{q}})}{2}(f_{++}+f_{--}) - \frac{1-\cos(\theta_{\mathbf{k}}+\theta_{\mathbf{k}+\mathbf{q}})}{2}(f_{+-}+f_{-+}).\end{aligned} \tag{S44}$$

Note that, at $\mathbf{q}=0$, we recover the result in Ref. [4] for the ferromagnetic and altermagnetic susceptibilities.

Using the previous results, we can evaluate the difference between the two susceptibilities,

$$\chi_0^{\mathrm{AM}}(\mathbf{q}, iq_n) - \chi_0^{\mathrm{FM}}(\mathbf{q}, iq_n) = \sum_{\mathbf{k}} \frac{t_{x,\mathbf{k}}t_{x,\mathbf{k}+\mathbf{q}}}{E^+_{\mathbf{k}}E^+_{\mathbf{k}+\mathbf{q}}}[f_{++}+f_{--}-(f_{+-}+f_{-+})]. \tag{S45}$$

This corresponds to Eq. (14) in the main text. In the case of $xz/yz$ orbitals, the $t_{x,\mathbf{k}}$ hopping is smaller compared to $xy$ orbitals. As a consequence, the altermagnetic and ferromagnetic susceptibilities shown in the main text are nearly degenerate.

## S5. RPA SUSCEPTIBILITIES

In the main text, we show the bare susceptibilities for the $xy$ and $xz/yz$ orbitals, using the expression derived in the previous section. Here, we use the random phase approximation (RPA) to examine the leading instabilities for the two uncoupled models, considering the intra-orbital Hubbard interaction

$$H_{\mathrm{intra}} = U\sum_{i,\mu} n_{i,\mu,\uparrow}n_{i,\mu,\downarrow}. \tag{S46}$$

For a given orbital $\mu$, the multi-orbital RPA susceptibility matrix in sublattice space is given by [4–8]

$$[\chi_{\mathrm{RPA}}(\mathbf{q}, iq_n)]^{i_1,i_2}_{i_3,i_4} = \left[\chi_0(\mathbf{q}, iq_n)\big(\mathbb{1}-U\chi_0(\mathbf{q}, iq_n)\big)^{-1}\right]^{i_1,i_2}_{i_3,i_4}, \tag{S47}$$

where $iq_n$ is a bosonic Matsubara frequency, $[U]^{i_1,i_2}_{i_3,i_4} = U$ for $i_1=i_2=i_3=i_4$ and the bare susceptibility matrix is given by

$$[\chi_0(\mathbf{q}, iq_n)]^{i_1,i_2}_{i_3,i_4} = -\frac{1}{N\beta}\sum_{\mathbf{k},i\omega_n} G^0_{i_1i_3}(\mathbf{k}+\mathbf{q}, i\omega_n+iq_n)G^0_{i_4i_2}(\mathbf{k}, i\omega_n). \tag{S48}$$

with the bare Green's function corresponding to

$$G^0_{ij}(\mathbf{k}, i\omega_n) = \sum_m \frac{u^i_m(\mathbf{k})u^{j*}_m(\mathbf{k})}{i\omega_n - E_m(\mathbf{k})}, \tag{S49}$$

with $u^i_m(\mathbf{k})$ the $\mathbf{k}$-dependent eigenvector connecting band space ($m$) with sublattice space ($i$), and $E_m(\mathbf{k})$ the corresponding energy eigenvalue of the Hamiltonian $H'$ for band $m$.

The RPA susceptibility in the ferromagnetic (FM) and altermagnetic (AM) channels are given by [4]

$$\chi^{\mathrm{FM}}_{\mathrm{RPA}}(\mathbf{q}, \omega) = \sum_{i,j}[\chi_{\mathrm{RPA}}(\mathbf{q}, iq_n\to\omega+i\eta)]^{i,i}_{j,j}, \tag{S50}$$

$$\chi^{\mathrm{AM}}_{\mathrm{RPA}}(\mathbf{q}, \omega) = \sum_{i,j}(-1)^{\mu}(-1)^{\nu}[\chi_{\mathrm{RPA}}(\mathbf{q}, iq_n\to\omega+i\eta)]^{i,i}_{j,j}. \tag{S51}$$

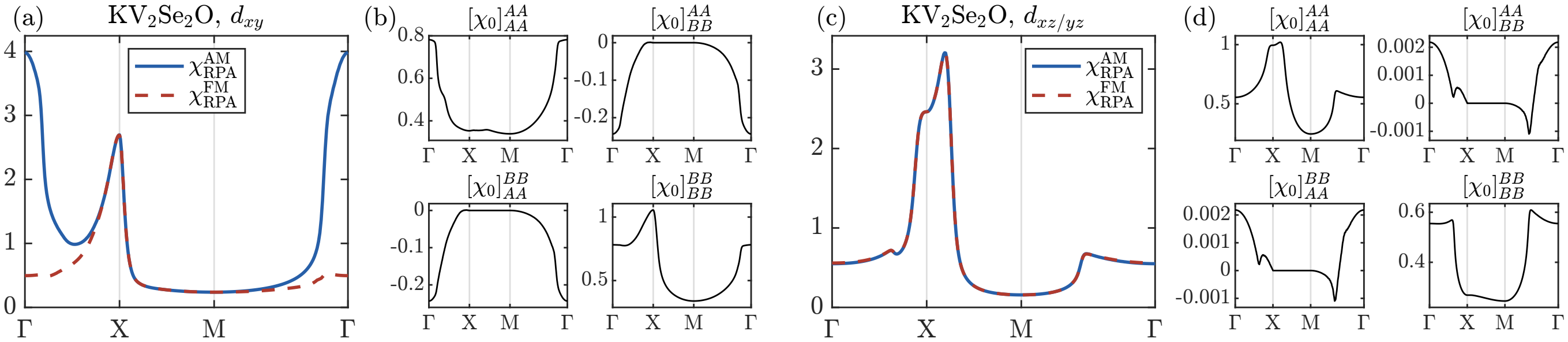


Figure S1. RPA susceptibility for the $xy$ (a) and $xz/yz$ orbitals (b) for $KV_2Se_2O$, using the parameters detailed in Table 1 of the main text. Panels (b) and (d) show the matrix elements of the respective bare susceptibilities.

If the altermagnetic susceptibility diverges before the ferromagnetic channel, it indicates that the altermagnetic phase is favored over ferromagnetism or spin-density wave order.

The RPA susceptibilities for $KV_2Se_2O$ are shown in Fig. S1. We note that the matrix elements of $[\chi_0]^{AA}_{AA}$ and $[\chi_0]^{BB}_{BB}$ are not identical but related by a $C_4$ rotation due to the $\tau_z$ term in the Hamiltonian. As a result, the FM and AM (Néel) channel of the bare susceptibility mix on the RPA level. This mixing is strongest at the X and Y points, where $|t_z|$ is maximal leading to a possible sublattice-and-direction selective divergence of the RPA susceptibility. For example, in the case of the $xz/yz$ orbital shown in Fig. S1(c), the divergence at the X (Y) point stems from the matrix element $[\chi_0]^{AA}_{AA}$ ($[\chi_0]^{BB}_{BB}$) seen in Fig. S1(d). As previously noted, based on the bare susceptibilities shown in Fig. 2 of the main text, the $xy$ orbitals drive the altermagnetic transition, as evident from the peak at Γ in the AM channel, despite the secondary peak at X being increased from the RPA due to the mixing of the bare AM and FM channels.

## S6. DERIVATION OF THE MULTIORBITAL MEAN FIELD HAMILTONIAN

As established in the main text, the interacting Lieb lattice Hamiltonian including on-site intra-orbital Coulomb repulsion ($U$), inter-orbital Coulomb repulsion ($U'$), Hund's coupling ($J_{\mathrm{H}}$) and pair hopping ($J'$) is

$$H = H_0 + U\sum_{i,\mu} n_{i\mu\uparrow}n_{i\mu\downarrow} + \sum_{i,\mu\neq\mu'}\left[\frac{1}{2}\sum_{\sigma\sigma'}\left(U' c^\dagger_{i\mu'\sigma'}c^\dagger_{i\mu\sigma}c_{i\mu\sigma}c_{i\mu'\sigma'} - J_{\mathrm{H}} c^\dagger_{i\mu'\sigma'}c^\dagger_{i\mu\sigma}c_{i\mu\sigma'}c_{i\mu'\sigma}\right) + J' c^\dagger_{i\mu\uparrow}c^\dagger_{i\mu\downarrow}c_{i\mu'\downarrow}c_{i\mu'\uparrow}\right], \quad \text{(S52)}$$

where $\mu\neq\mu'$ denotes different orbitals and we use the common approximation of spherical symmetric $d$-shell orbitals, i.e. $U' = U - 2J_H$ and $J' = J_H$. We expand in same orbital mean fields $\langle c^\dagger_{i\mu\sigma}c_{i\mu\sigma'}\rangle$ and define the spin operator expectation values $\langle M_{s,i\mu}\rangle = \langle \Psi^\dagger_{i\mu}\sigma_s\Psi_{i\mu}\rangle$ and the density $\langle n_{i\mu}\rangle = \langle \Psi^\dagger_{i\mu}\Psi_{i\mu}\rangle$ for each orbital $\mu$, where we use the basis $\Psi_{i\mu} = (c_{i\mu\uparrow}, c_{i\mu\downarrow})^T$ and $\sigma_s$, $s\in\{x,y,z\}$ are the spin Pauli matrices. From now on, we assume a homogeneous system, and the site index $i$ becomes a sublattice index $i\in\{A,B\}$, keeping in mind that we only deal with local interactions that become constant in momentum space. In this study, the spin is rotationally invariant in the absence of SOC or aligned with the direction of SOC and we choose the spin basis such that only $\sigma_z$ has non-zero expectation values. We note that the Néel order parameter for altermagnetic order $N_{z,\mu}\propto\langle M_{z,A\mu}\rangle - \langle M_{z,B\mu}\rangle$ and the macroscopic magnetization $M_{z,\mu}\propto\langle M_{z,A\mu}\rangle + \langle M_{z,B\mu}\rangle$. We obtain the mean field Hamiltonian

$$H_{\mathrm{MF}} = H_0 + \sum_{i,\mu}\left[\frac{U}{2}\left(\langle n_{i\mu}\rangle n_{i\mu} - \langle M_{z,i\mu}\rangle M_{z,i\mu}\right)\right] + \sum_{i,\mu\neq\mu'}\left[\left(U' - \frac{J_H}{2}\right)\langle n_{i\mu}\rangle n_{i\mu'} - \frac{J_H}{2}\langle M_{z,i\mu}\rangle M_{z,i\mu'}\right] \quad \text{(S53)}$$

that needs to be solved self consistently.

We calculate the grand canonical potential

$$\Omega = \sum_{m\mu\sigma\mathbf{k}} E_{m\mu\sigma\mathbf{k}} f(E_{m\mu\sigma\mathbf{k}}) - TS - E[n_{i\mu}, M_{z,i\mu}] \quad \text{(S54)}$$

with temperature $T$ and entropy

$$S = -k_{\mathrm{B}}\sum_{m\mu\sigma\mathbf{k}}\left[f(E_{m\mu\sigma\mathbf{k}})\ln(f(E_{m\mu\sigma\mathbf{k}})) + (1 - f(E_{m\mu\sigma\mathbf{k}}))\ln(1 - f(E_{m\mu\sigma\mathbf{k}}))\right]. \quad \text{(S55)}$$

Here, $E_{m\mu\sigma\mathbf{k}}$ are the eigenenergies of $H_{\rm MF}$ with band index $m$ for each orbital and spin and

$$E[n_{i\mu}, M_{z,i\mu}] = \frac{U}{4}\left(\langle n_{i\mu}\rangle^2 - \langle M_{z,i\mu}\rangle^2\right) + \left(\frac{2U' - J_H}{4}\right)\langle n_{i\mu}\rangle\langle n_{i\mu'}\rangle - \frac{J_H}{4}\langle M_{z,i\mu}\rangle\langle M_{z,i\mu'}\rangle \tag{S56}$$

is the constant energy contribution resulting from the mean field expectation values that is relevant when comparing free energies. Since we perform calculations at fixed total particle number $\langle n\rangle = \sum_{\sigma\mu} n_{\sigma\mu}$ and separately fixed filling $\langle n_\mu\rangle = \langle n_{\uparrow\mu} + n_{\downarrow\mu}\rangle$ for each of the orbtials, we enforce these via Lagrange multipliers during the selfconsistent calculation to minimize the Helmholtz free energy

$$F = \Omega + \tilde{\mu}\langle n\rangle + \Delta(\langle n_{xy}\rangle - \langle n_{xz/yz}\rangle). \tag{S57}$$

We perform selfconsistent calculations where the system is initialized in an AM state ($\langle\tau_z\sigma_z\rangle \neq 0$) or a FM state ($\langle\tau_0\sigma_z\rangle \neq 0$) and compare the free energies of the converged solutions to determine the ground state, as shown in Fig. 3 of the main text.

### A. Single-orbital mean field calculations

We study the effect of varying the orbital filling $n$ and the strength of the altermagnetic hopping anisotropy on the orbitals' free energy contribution, and the evolution of the order parameter for the single-orbital models discussed in the main text. We quantify the strength of the hopping anisotropy by the parameter $\alpha$ that scales all hopping terms $t_{z,i}$, where $\alpha = 0$ corresponds to an effective antiferromagnet and $\alpha = 1$ is the value used in the main text. Here, we focus on the case where the orbitals are not Hunds-coupled ($J_H = 0$) to extract inherent contributions and ordering tendencies for the respective orbitals.

Figure S2 shows the tendency of the $xy$ orbitals in KVSO to form Néel order across all studied parameters (panels a,b), apparent by the significantly lower free energy. In contrast, the altermagnetic and ferromagnetic state have a similar condensation energy for the $xz/yz$ orbitals, with the ferromagnetic state being the ground state by a small margin. This is consistent with our findings from the RPA susceptibility calculations, see Fig. S1. The same observation can be made for RVTO in Fig. S3. The difference is that the onset of magnetic order for the $xz/yz$ orbitals is at a lower value of $U$, explaining the region where FM order is stabilized for 1.4 eV $\leq U \leq$ 1.8 eV (Fig. 3(b) in the main text) before the $xy$ orbitals become magnetic as well, and Néel order becomes favorable again even at $J_H = 0$. As shown in Fig. S4 (c,d), the main effect of spin orbit coupling in KVSO is that the energy difference between the FM and AM state in the $xz/yz$ orbitals becomes larger, increasing the region of FM in the phase diagram shown in Fig. 4 of the main text.

## S7. ORBITAL-SELECTIVE TOPOLOGICAL EDGE STATES

We examine the stability of our HF calculation in the presence of relativistic SOC for $KV_2Se_2O$ (see the parameter set in Table I of the main text). This model hosts topological states for $|N/2t_{z,1}| < 1$. As seen in Fig. S5, SOC enhances the region where ferromagnetism is favorable for weak Hund's coupling. Yet, at moderate $J_H$, the altermagnetic state is readily stabilized. Panels (c,f) demonstrate the altermagnetic order parameter being enhanced by Hund's coupling, whereas it is barely affected by the coupling in the ferromagnetic case. Notably, $M_{xy} = U/2\,\langle\tau_0\sigma_z\rangle$ is weak throughout the studied parameter regime, and $N_{xy} = U/2\,\langle\tau_z\sigma_z\rangle$ drives the altermagnetism. Panels (a,b) show that the transition from the topological to the non-topological regime takes place at different interaction strengths for the two models, revealing a regime in which the topological states are orbital selective.

## S8. SOC COUPLING THE $xy$ AND $xz/yz$ TIGHT-BINDING MODELS

As discussed in the main text, the single-particle Hamiltonians for $xy$ and $xz/yz$ orbitals are uncoupled since the hopping is not allowed due to the mirror symmetry $\{m_z|\mathbf{0}\}$. Therefore, the two models are only coupled through electron interactions and SOC. In the main text, we focus on the multi-orbital electronic interactions, which is expected to have the largest effect. For completeness, here we detail the form of the momentum-independent SOC coupling the two models.

We choose the basis in orbital and sublattice space as $\{d_{xy,A}, d_{xy,B}, d_{xz,A}, d_{yz,B}\}$. Therefore, in this basis, the non-interacting Hamiltonian can be written as

$$H_0 = \frac{\gamma_0 + \gamma_3}{2} H_0^{xy} + \frac{\gamma_0 - \gamma_3}{2} H_0^{xz/yz} + H_{\rm SOC}, \tag{S58}$$

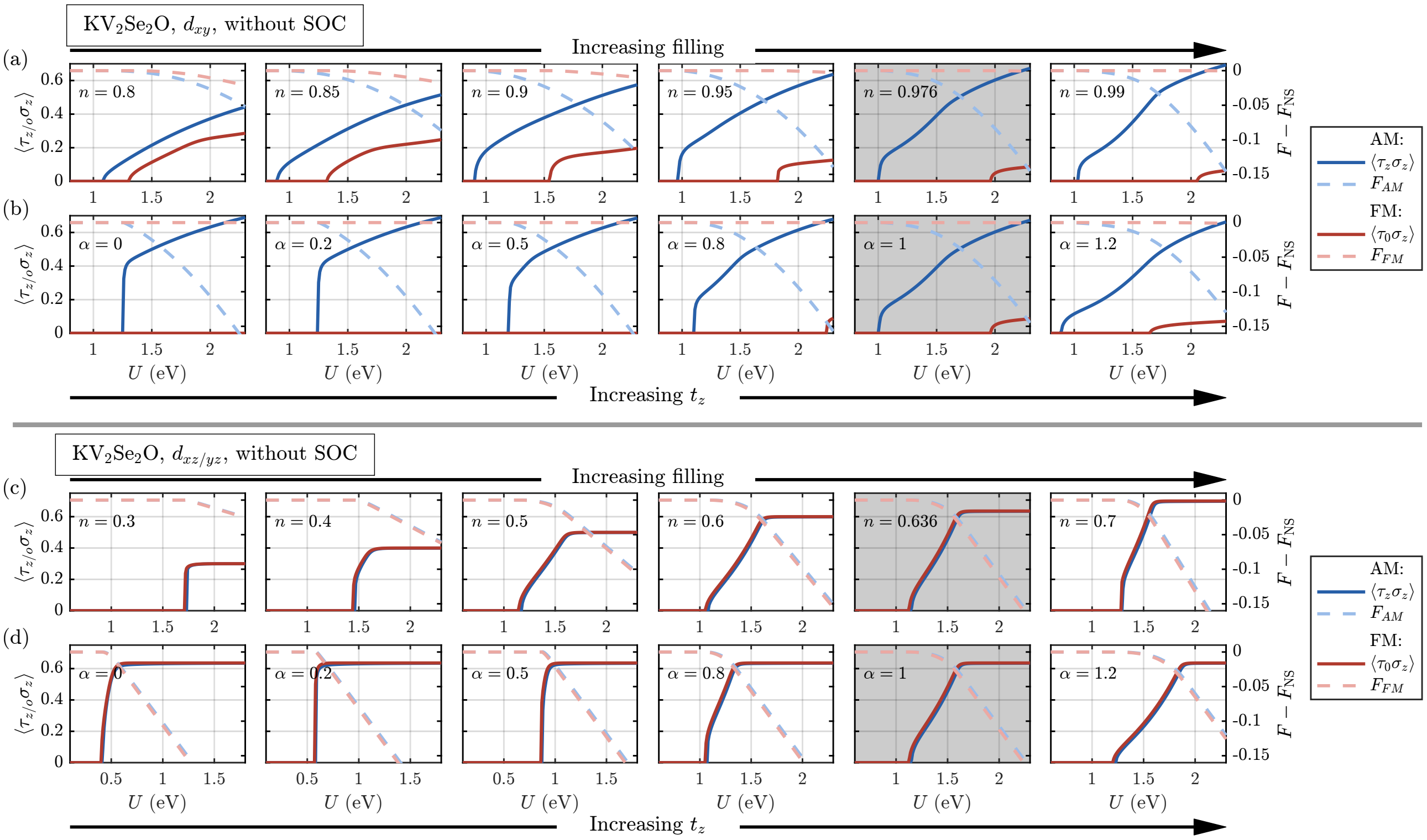


Figure S2. HF mean field results for the KVSO $d_{xy}$ orbital (a,b) and the $d_{xz/yz}$ orbital (c,d) for varying filling $n$ (a,c) and varying altermagnetic hopping $t_z = \alpha t_{z,0}$. Shown are the AM (blue) and FM (red) order parameter (full lines, left axis) and Helmholtz free energy (dashed lines, right axis). The gray highlighted panel corresponds to the $J_H = 0$ case in the main text.

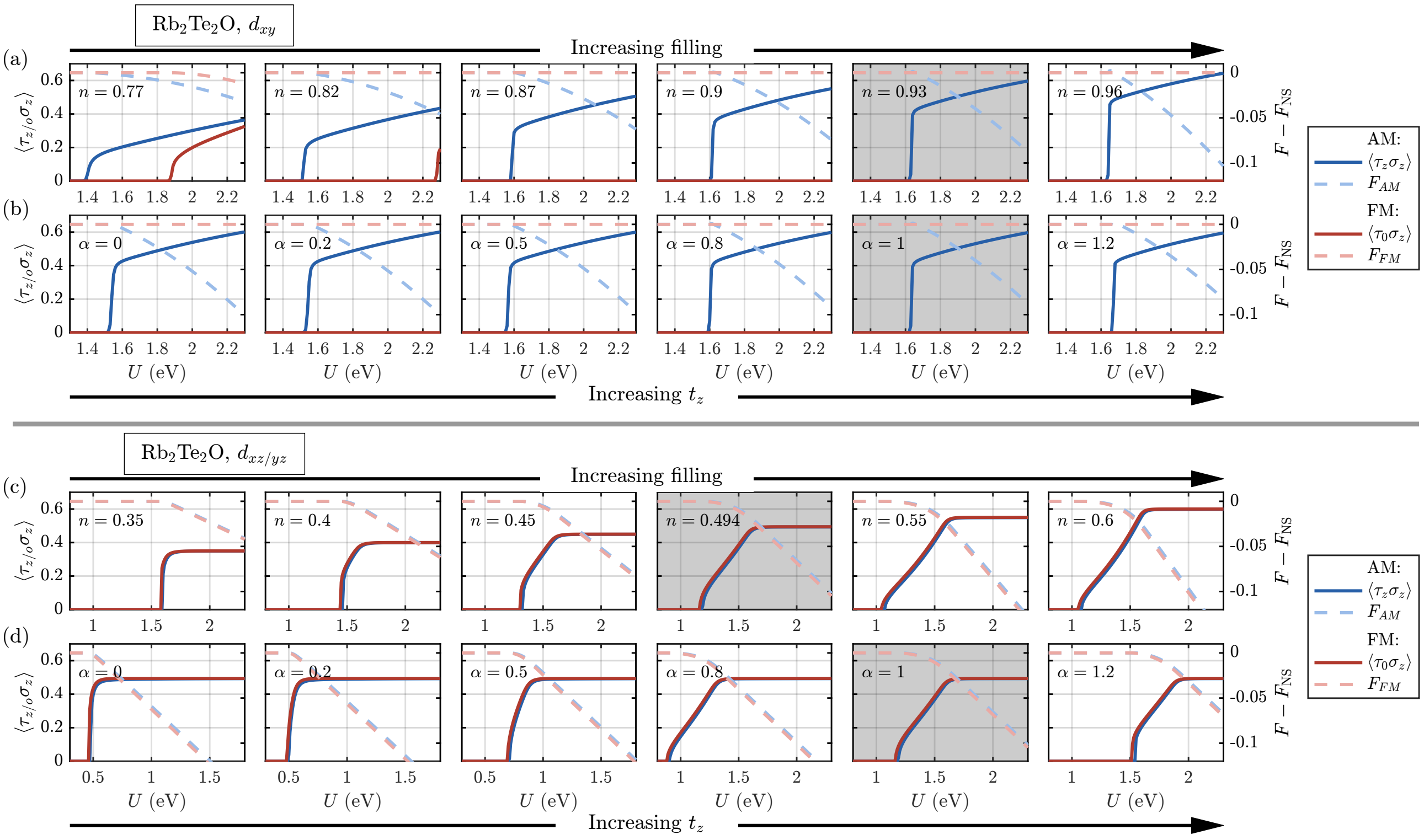


Figure S3. HF mean field results for the RVTO $d_{xy}$ orbital (a,b) and the $d_{xz/yz}$ orbital (c,d) for varying filling $n$ (a,c) and varying altermagnetic hopping $t_z = \alpha t_{z,0}$. Shown are the AM (blue) and FM (red) order parameter (full lines, left axis) and Helmholtz free energy (dashed lines, right axis). The gray highlighted panel corresponds to the $J_H = 0$ case in the main text.

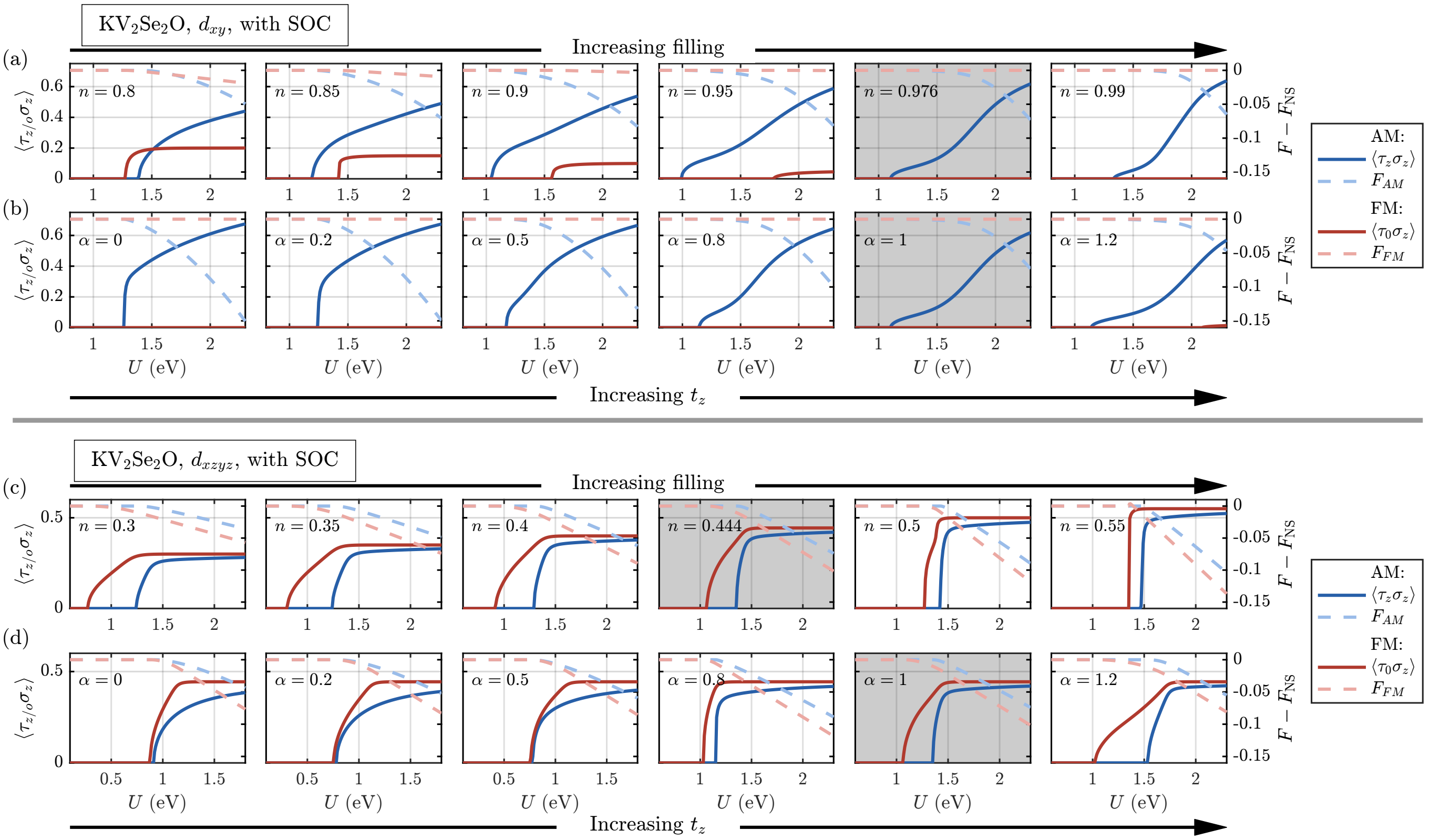


Figure S4. Hartree-Fock mean field results for the RVTO $d_{xy}$ orbital (a,b) and the $d_{xz/yz}$ orbital (c,d) for varying filling $n$ (a,c) and varying altermagnetic hopping anisotropy $t_z = \alpha t_{z,0}$, in the presence of SOC. Shown are the AM (blue) and FM (red) order parameter (full lines, left axis) and Helmholtz free energy (dashed lines, right axis). The gray highlighted panel corresponds to the $J_H = 0$ case in the end matter.

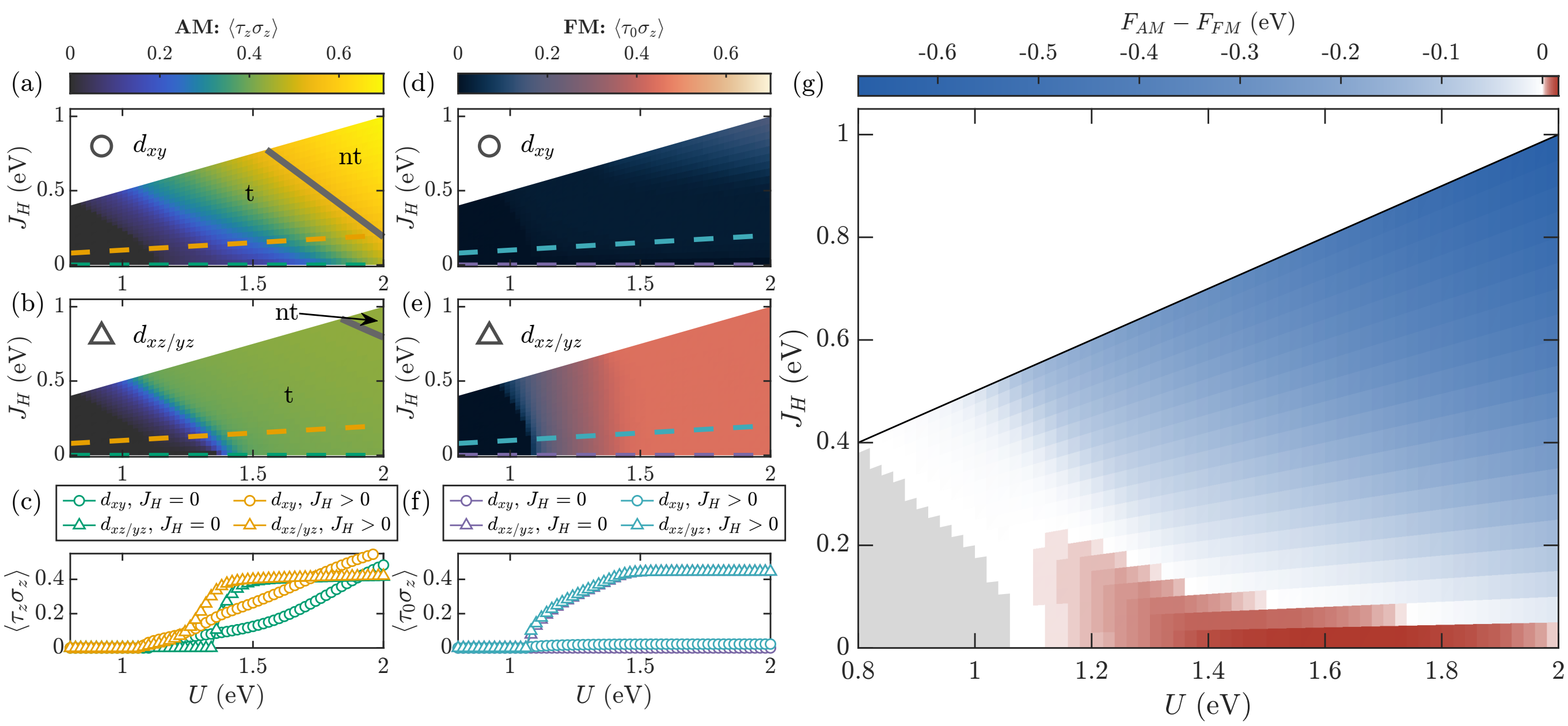


Figure S5. Order parameter of the AM (a-c) and FM (d-f) state for $KV_2Se_2O$ with SOC in $U$–$J_H$ parameter space. (a,d) and (b,e) show the $d_{xy}$ and $d_{xz/yz}$ order parameters, respectively. (c,f) show cuts for zero and finite values of $J_H$ as indicated by the lines in the phase diagrams. We mark the topological (t) and non-topological (nt) regimes for the orbitals in the altermagnetic state. (g) Helmholtz free energy difference showing regions of the phase diagram where AM (blue) and FM (red) are favorable. The gray region highlights the non-magnetic state.

where $\gamma_i$ are the Pauli matrices in orbital space and the single-particle Hamiltonians for $xy$ and $xz/yz$ orbitals are introduced in Eqs. (1)-(5) of the main text. In the equation above, the two models are coupled by the SOC Hamiltonian $H_{\rm SOC}$, which is given by

$$H_{\rm SOC} = \begin{pmatrix} 0 & 0 & -i\sigma_x & 0 \\ 0 & 0 & 0 & i\sigma_y \\ i\sigma_x & 0 & 0 & 0 \\ 0 & -i\sigma_y & 0 & 0 \end{pmatrix}. \tag{S59}$$

Hence, there is a momentum-independent SOC that couples the two tight-binding Hamiltonians.


\fi
\end{document}